\documentclass{aa}
\usepackage{graphicx}
\usepackage{xcolor}
\usepackage{amsmath}
\usepackage{float}
\usepackage{gensymb}
\usepackage[bookmarks, colorlinks, breaklinks]{hyperref} 
\hypersetup{linkcolor=blue,citecolor=blue,filecolor=blue,urlcolor=blue} 
\usepackage{txfonts}

\def\btheta{\mbox{\boldmath $\theta$}}

\def\ge{\geqslant}

\def\sex{{\tt SExtractor}}
\def\psfex{{\tt PSFEx}}
\def\scamp{{\tt scamp}}
\def\swarp{{\tt SWarp}}

\def\astromatic{Astr\textit{O}matic}

\newcommand{\pnt}{P$_{\text{nt}}$}
\newcommand{\ptot}{P$_{\text{tot}}$}
\newcommand{\pth}{P$_{\text{th}}$}
\newcommand{\rfive}{R$_{500c}$}
\newcommand{\rtwoc}{R$_{200c}$}

\newcommand{\mtwoc}{M$_{200c}$}
\newcommand{\ctwoc}{$c_{200c}$}
\newcommand{\msun}{M$_{\odot}$}
\newcommand{\xmm}{XMM-{\it Newton}}
\newcommand{\planck}{{\it Planck}}
\newcommand{\hst}{{\it HST}}

\def\cluster{PSZ2 G313.33+61.13 }

\begin{document}

   \title{CHEX-MATE: Cluster Multi-Probes in Three Dimensions (CLUMP-3D)}
   \subtitle{II. Combined Gas and Dark Matter Analysis from X-ray, SZE, and WL}

   \author{A. Gavidia\inst{\ref{caltech}}
          \and
          J. Kim\inst{\ref{kaist}}
          \and
          J. Sayers\inst{\ref{caltech}}
          \and 
          M. Sereno\inst{\ref{inaf-oas}, \ref{infn-bo}}
          \and
          L. Chappuis\inst{\ref{cea}, \ref{gva}}
          \and
          D. Eckert\inst{\ref{gva}}
          \and
          K. Umetsu\inst{\ref{asiaa}}
          \and
          H. Bourdin\inst{\ref{tv}, \ref{infnro}}
          \and 
          F. De Luca\inst{\ref{tv}, \ref{infnro}}
          \and
          S. Ettori\inst{\ref{inaf-oas}, \ref{infn-bo}}
          \and
          M. Gaspari\inst{\ref{umod}}
          \and
          R. Gavazzi\inst{\ref{inst1}, \ref{inst2}}
          \and
          S. Kay\inst{\ref{jbc}}
          \and 
          L. Lovisari\inst{\ref{inaf-mil}, \ref{cfa}}
          \and
          P. Mazzotta\inst{\ref{tv}, \ref{infnro}}
          \and
          G. W. Pratt\inst{\ref{cea}}
          \and
          E. Rasia\inst{\ref{inaf-tri}, \ref{umich}}
          \and
          M. Rossetti\inst{\ref{inaf-mil}}
          }

   \institute{California Institute of Technology, 1200 E. California Blvd., MC 367-17, Pasadena, CA 91125, USA\label{caltech} \\
              \email{agavidia@caltech.edu} 
              \and
              Korea Advanced Institute of Science and Technology (KAIST), 291 Daehak-ro, Yuseong-gu, Daejeon, 34141, Republic of Korea\label{kaist}
              \and
              INAF - Osservatorio di Astrofisica e Scienza dello Spazio di Bologna, via Piero Gobetti 93/3, I-40129 Bologna, Italy \label{inaf-oas}
              \and
              INFN, Sezione di Bologna, viale Berti Pichat 6/2, I-40127 Bologna, Italy \label{infn-bo}
              \and 
              Université Paris Saclay, IRFU/CEA, 91191, Gif-sur-Yvette, France\label{cea}
              \and 
              Department of Astronomy, University of Geneva, Ch. d’Ecogia 16, CH-1290 Versoix, Switzerland\label{gva}
              \and
              Academia Sinica Institute of Astronomy and Astrophysics (ASIAA), No. 1, Section 4, Roosevelt Road, Taipei 106216, Taiwan \label{asiaa}
              \and
              Dipartimento di Fisica, Universit\`a degli studi di Roma Tor Vergata, Via della Ricerca Scientifica 1, I-00133 Roma, Italy \label{tv}
              \and 
              INFN, Sezione di Roma ‘Tor Vergata’, Via della Ricerca Scientifica, 1, 00133, Roma, Italy \label{infnro}
              \and
              Department of Physics, Informatics and Mathematics, University of Modena and Reggio Emilia, 41125 Modena, Italy \label{umod}
              \and
              Laboratoire d’Astrophysique de Marseille, Aix-Marseille Univ., CNRS, CNES, 13013 Marseille, France \label{inst1} 
              \and 
              Institut d’Astrophysique de Paris, UMR7095 CNRS \& Sorbonne Université, 98bis Bd Arago, 75014, Paris, France \label{inst2}
              \and 
              Jodrell Bank Centre for Astrophysics, Department of Physics and Astronomy, The University of Manchester, Manchester M13 9PL, UK \label{jbc}
              \and 
              INAF, Istituto di Astrofisica Spaziale e Fisica Cosmica di Milano, via A. Corti 12, 20133 Milano, Italy \label{inaf-mil}
              \and
              Center for Astrophysics $|$ Harvard $\&$ Smithsonian, 60 Garden Street, Cambridge, MA 02138, USA \label{cfa}
              \and 
              INAF – Osservatorio Astronomico di Trieste, via Tiepolo 11, I-34131 Trieste, Italy \label{inaf-tri}
              \and 
              Department of Physics; University of Michigan, Ann Arbor, MI 48109, USA \label{umich}}

   \date{}

  \abstract
  {
  Under the standard model of hierarchical structure formation, the overall geometry of galaxy clusters is better described by a triaxial ellipse than a sphere. 
  As a result, applying spherically-symmetric models can result in significant biases, with masses derived from WL observations being particularly sensitive. These biases can be mitigated by fitting a triaxial model, but this requires deep multiprobe data along with a set of physically motivated models to describe them.
  Here we present a multiprobe triaxial analysis methodology based on the data available for galaxy clusters in the Cluster Heritage project with \xmm\ - Mass Assembly and Thermodynamics at Endpoint of structure formation (CHEX-MATE),
  which includes X-ray data from \xmm, SZ data from \planck\ and ACT, and WL data from Subaru. This work builds upon our previous development of a gas-only X-ray and SZ triaxial fitting formalism in Paper I. After verifying our approach using mock observations of model clusters with known properties, we apply it to the CHEX-MATE galaxy cluster \cluster (Abell 1689). We find that the cluster is elongated along the line of sight relative to the plane of sky by a factor of $\mathcal{R}_{LP} = 1.27 \pm 0.02$. As a result, the WL mass obtained from our triaxial fit, \mtwoc\ $=(13.69_{-1.41}^{+1.56})\times10^{14}$ \msun, is significantly lower than the value of $(17.77_{-1.75}^{+2.00})\times10^{14}$ \msun\ obtained from a spherically-symmetric fit that otherwise employs the same methodology. Our triaxial fit finds a concentration of \ctwoc$=8.55_{-1.61}^{+2.20}$, consistent with the spherically-symmetric value of $9.99_{-1.78}^{+2.26}$, which suggests that the unexpectedly high concentration in Abell 1689 is not due to triaxiality and orientation. We also measure the non-thermal pressure fraction at radii between 0.18--1.37~Mpc, finding a minimum of approximately 20~per cent at intermediate radii increasing to near 30~per cent at both the smallest and largest radii, and with a typical measurement precision of $\pm 5$~per cent.  
}

   \keywords{Galaxies: clusters: individual: Abell 1689 --- Galaxies: clusters: general --- X-rays: galaxies: clusters --- Galaxies: clusters: intracluster medium --- Gravitational lensing: weak --- Cosmology: observations}  

\maketitle
%

\section{Introduction}
Within the current standard cosmological framework, structure formation proceeds as a hierarchical merging process \citep{Davis1985, Kravtsov2012}, with galaxy clusters forming from the gravitational collapse of the most dense peaks in the initial primordial density field \citep{Kaiser1986, Kaiser1991}. Anisotropic collapse results in a web of filaments, which connect galaxy clusters and other large objects and continue to feed their growth \citep{Bond1996}. Within this geometry, the primary nodes of the web (i.e., clusters) are better described by triaxial ellipses rather than spheres \citep{Limousin2013,Despali2014,Bonamigo2015}, with numerical simulations predicting preferential alignment between the elongated axes and the large scale connected filaments \citep{Frenk1988, Dubinski1991, Warren1992, Cole1996, Jing2002, Hopkins2005, Bailin2005, Kasun2005, Paz2006a, Allgood2006, Bett2007, Munoz2010, Gao2012}.

A range of observational evidence demonstrates that clusters are elliptical in the two-dimensional plane of sky (POS). In particular, the projected signals from various probes, such as X-ray surface brightness \citep[SB,][]{Lau2012}, the Sunyaev-Zeldovich (SZ) effect scattering of cosmic microwave background photons \citep{Sayers2011a}, and both strong and weak gravitational lensing \citep[SL and WL,][]{Oguri2010} generally display some degree of elongation in the POS. While clusters have long been known to deviate from spherical symmetry, limitations in data quality and availability have often resulted in modeling them as spherical objects, particularly for SZ \citep[e.g.,][]{Bleem2024} and WL \citep[e.g.,][]{Chiu2025} observations. Even with exquisite data, because the observed signals are a two-dimensional projection, critical three-dimensional geometric information is lost, generally preventing analyses employing a single observational probe from fitting non-spherical models. To accurately recover triaxial cluster morphology, one must combine deep multi-probe observational data \citep[e.g.][]{Limousin2013}. 

For example, the three-dimensional properties of a cluster can be inferred from the combination of X-ray and SZ data by utilizing their respective dependencies on the intra-cluster medium (ICM) gas density and temperature. These data redundantly probe ICM thermodynamics, but with different weighting for the line of sight (LOS) projection of the observed signals. The X-ray emission is primarily from thermal Bremsstrahlung \citep{Sarazin1986}, with 
\begin{displaymath}
    S_\text{X} \propto \int n_e^2 \Lambda(T_e, Z) dl,
\end{displaymath}
where $n_e$ is the electron density and $\Lambda(T_e, Z)$ is the cooling function which depends on temperature ($T_e$) and metallicity ($Z$). For typical ICM temperatures and metallicities, $\Lambda(T_e, Z)$ is approximately constant for soft-band measurements of $S_{\text{X}}$ at $\lesssim 2$~keV \citep{Ettori2000}. In contrast, the SZ signal brightness \citep{Sunyaev1972} is given by
\begin{displaymath}
    \centering
    B_{\text{SZ}} \propto \int n_e T_e dl.
\end{displaymath}
Spectroscopic fits to X-ray observations can also measure the ICM temperature, $T_\text{X}$, and so in the limit of an isothermal and isodensity gas distribution along the LOS, the extent of the cluster along the LOS, $\Delta l$, can be obtained through the combination of these three measurements as 
\begin{displaymath}
    \Delta l \sim \frac{B_{\text{SZ}}^2 \Lambda(T_e, Z)}{S_\text{X} T_\text{X}^2} \sim \frac{B_{\text{SZ}}^2}{S_\text{X} T_\text{X}^2},
\end{displaymath} 
where the right hand expression is approximately valid for $S_\text{X}$ measured in the soft band. In practice, deviations from isothermality and isodensity prevent such a simplistic calculation, but information on the LOS extent of the cluster can still be obtained from jointly modeling the X-ray and SZ data by utilizing elliptically-symmetric parametric models to describe the spatial variations in these thermodynamic quantities. An early analysis based on this approach was performed by \citet{DeFilippis2005}, who found the spherical hypothesis was strongly rejected for the majority of objects in a sample of 25 clusters. In a later study considering the same data, \citet{Sereno2006} found a mixed population of prolate and oblate morphologies, with prolate shapes preferred among 60-76~per cent of the sample. Subsequent triaxial analyses based on this general technique have all found measurable deviations from spherical symmetry \citep[e.g.,][]{Morandi2011, Morandi2012_A383, Morandi2012_A1835, Sereno2017, Sereno2018, Kim24}

Alternatively, SL and WL datasets can be combined to make de-projected surface density maps, and to then determine the triaxial structure of the cluster from them. This method typically assumes a Navarro-Frenk and White \citep[NFW,][]{Navarro1997} density profile in an elliptical coordinate basis. The projected NFW profile can be described in terms of the strength of the lens and the projected length scale, which is a function of cluster shape and orientation, and is fit to the observed lensing signals (i.e., shear, convergence, magnification). This approach has been applied to single clusters \citep[e.g.,][who find mildly triaxial halos]{Sereno2011, Umetsu2012}, and to an analysis of 20 clusters from the Custer Lensing And Supernovae survey with Hubble \citep[CLASH,][]{Postman2012} sample, with \citet{Chiu2018} finding a preference for prolate geometry in good agreement with the results from the joint X-ray and SZ analyses detailed above.

This work builds upon the triaxial fitting pipeline described in \citet[][hearafter K24]{Kim24}, which is based on the combination of SZ and X-ray data to obtain triaxial geometry, by adding a WL analysis to their gas-only approach, while additionally building on the WL fitting methodology described in \citet{Sereno2017}. This expansion of the pipeline allows for a measurement of the total cluster mass. In more detail, K24 performs a joint X-ray and SZ analysis to constrain the geometrical properties of a cluster, which includes two axial ratios that give the relative lengths of the three orthogonal primary axes of a triaxial ellipse, along with their orientation relative to the LOS and POS. \citet{Sereno2017} includes a total mass profile in their model which is constrained by the combination of SL and WL. They assume constant axial ratios for this profile, which are different from the gas axial ratios, but coaligned. As detailed below in Section.~\ref{sec:mass_reconstruction}, we instead utilize only WL data and we assume constant axial ratios for the gravitational potential, which are equal to the gas axial ratios. This decision is motivated by the potential being smoother that the total mass profile \citep{Stapelburg2022}.

As a particular motivation to add WL to the model of K24, we note that the three-dimensional ellipticity of clusters, combined with orientation, is the primary cause of bias in single-cluster mass measurements from WL \citep[e.g.,][]{Meneghetti2010,Giocoli2024}. This is due to the especially strong degeneracy between total mass and elongation along the LOS, and fitting a three-dimensional elliptical model allows most of this bias to be eliminated. In addition, cluster orientation can impact the inferred shape of the radial profile of mass density, which is often quantified by a mass-dependent concentration in the context of the NFW model. Past studies have shown that exceptionally large concentrations found for individual clusters were a result of elongation \citep{Broadhurst2008, Oguri2009, Sereno2010, Meneghetti2014}. Thus, a WL analysis that includes three-dimensional shape information provided by the gas observables offers the promise of minimally biased individual cluster mass and concentration measurements.

These relatively unbiased mass reconstructions also open the possibility to probe the level of support against gravitational collapse due to motions outside of hydrostatic equilibrium (HSE) in the ICM, often referred to as non-thermal pressure (\pnt). There are various contributions to \pnt\ in the central and outer regions of the cluster. In the cluster core, feedback from a centrally located active galactic nucleus (AGN) is likely the primary heating source \citep{McNamara2005}. This heating is dynamic, as AGN jets inflate bubbles in the gas and transfer energy via multiple mechanisms, such as shocks, cavity heating, and convective mixing. Additionally, turbulence is driven on smaller scales by core sloshing from cluster merger events, dynamical friction, and galaxy motions \citep{Rishi2016, Gaspari2020, Romero2025}. Direct measurements of the velocity structure in this region are enabled by X-ray spectroscopy from instruments like the X-Ray Imaging and Spectroscopy Mission \citep[XRISM,][]{Tashiro2025}. To date, XRISM has mainly focused on probing the region within the central $\sim$100 kpc in radius, extending beyond this region in one case, typically finding very low non-thermal pressure fractions \citep[\pnt/\ptot\ $\sim$2-3 per cent,][]{XRISM_A2029_part2, XRISM_A2029, XRISM_coma, XRISM_centaurus}. These results are, in general, lower than those found from hydrodynamical simulations of clusters. However, accurately simulating the centrally located AGN is an active research area, and so it is unclear if these XRISM measurements are in tension with the predicted values near 10 per cent found in such simulations \citep{Nelson2014, Angelinelli2020}. 

In this work, we focus on accurately reconstructing the distribution of gas and total mass, along with the thermodynamic properties of the gas, to obtain a measurement of \pnt\ in the outer cluster regions beyond 1 Mpc in radius. Numerical simulations suggest that clusters should progressively deviate more from HSE with increasing radius well beyond \rtwoc\ ($\sim$2-3 Mpc), due to newly accreted material which has not fully thermalized \citep{Piffaretti2008, Lau2009, Nelson2014, Angelinelli2020}. The in-falling material in this region is brought into thermal equilibrium with the ICM primarily through a series of radial shocks \citep{Miniati2000}, and the overall thermalization efficiency of these shocks can be quantified by the ratio of \pnt\ to the total pressure (\ptot). We note that \pnt\ includes both turbulent and coherent motions. Modeling \pnt\ in simulations of cluster outskirts can be challenging due to resolution effects related to the low particle density in these regions, and so it is currently unclear whether they fully trace the relevant physics. Simulations generally agree that \pnt/\ptot\ should increase with radius, but the predicted values of \pnt/\ptot\ near \rtwoc\ span a wide range from approximately 15--40~per cent \citep{Nelson2014, Shi2014, Shi2015, Vazza2018, Eckert2019, Pearce2020, Angelinelli2020, Sayers2021}. Observational studies of the ensemble average value of \pnt/\ptot\ from modest samples of $\simeq 10$ clusters also span a wide range, but have generally found lower values, with \pnt/\ptot\ $\simeq 10$--25~per cent near \rtwoc\ \citep{Siegel2018, Eckert2019, Sarkar2025}, potentially suggesting more efficient thermalization than typically found in the simulations.

In this paper, we demonstrate our triaxial modeling pipeline with a multi-probe dataset assembled as part of the Cluster HEritage project with \xmm: Mass Assembly and Thermodynamics at the Endpoint of Structure Formation collaboration\footnote{xmm-heritage.oas.inaf.it} \citep[CHEXMATE;][]{chexmate2021}, including SZ, X-ray SB and temperature, and WL shear measurements of the galaxy cluster \cluster (hereafter Abell 1689). The SZ and X-ray portion of the analysis largely follows that detailed in K24. We focus mainly on the integration of a WL analysis into the established gas-only analysis, and the subsequent mass profile and non-thermal pressure measurements obtained from this updated pipeline. In Section~\ref{sec:Section2} we summarize the triaxial analysis methodology, including our modeling formalism. In Section~\ref{sec:Pnt} we describe the calculation performed to obtain a radial profile of the non-thermal pressure support from our triaxial fits. Section~\ref{sec:MockData} reviews the results from running the fitting pipeline on a set of mock observations of known cluster models. This provides an understanding of potential biases introduced by the fit and demonstrates that we can accurately recover the known input parameters. We describe the observed data set for Abell 1689 in Section~\ref{sec:data}, and results of the fit to this data in Section~\ref{sec:Results}. A comparison to other results in the literature is detailed in Section~\ref{sec:Discussion}, including those obtained by \citet{Loris2025} using the same CHEX-MATE data but a much different modeling formalism. Throughout this study, we adopt a $\Lambda$CDM cosmology characterized by $H_0 = 70$km s$^{-1}$ Mpc$^{-1}$, $\Omega_M = 0.3$, and $\Omega_\Lambda = 0.7$.

\section{Triaxial Analysis Formalism} \label{sec:Section2}
\subsection{Geometry and Projection}
Our cluster model assumes a triaxial shape and we follow the mathematical description of a triaxial cluster geometry developed in K24, summarized here. The intrinsic cluster geometry is connected to the projected properties observed in the POS by writing the thermodynamic profiles of the ICM and the gravitational potential as functions of $\zeta$, the ellipsoidal radius. It is defined in the intrinsic coordinate system of the ellipsoid (x$_1$, x$_2$, x$_3$) as
\begin{equation}
    \zeta^2 = \frac{x_1^2}{q_1^2} + \frac{x_2^2}{q_2^2} + x_3^2,
    \label{eq:zeta}
\end{equation}
where $q_1$ is the minor-to-major axial ratio and $q_2$ is the intermediate-to-major axial ratio. The ellipsoid's intrinsic axes may not align with those of the observer, so we relate the two using three Euler angles: $\theta$, $\phi$, and $\psi$. Intrinsic parameters of the ellipsoid are its semi-major axis $l_p$, ellipticity $\epsilon$, and orientation $\theta_{\epsilon}$ in the POS, and the elongation parameter $\mathcal{R}_{LP}$, defined as the ratio of its LOS extent to the effective radius of its projected ellipse (i.e. its average POS extent). Note that we use a different definition for elongation than in K24, who define the parameter $e_\parallel$ as the ratio of the LOS extent and $l_p$. While $e_\parallel$ is a useful geometric quantity which will be used in the projection of the three-dimensional distributions (see Eq.~\ref{eq:Proj}), we use $\mathcal{R}_{LP}$ when describing the elongation of the cluster. 

The three-dimensional distributions are projected and expressed as functions of $\xi$, the two-dimensional projected elliptical radius, via 
\begin{equation}
    F_{2D}(x_\xi; l_p, p_i) = 2l_p e_{\parallel} \int_{x_\xi}^\infty F_{3D}(x_\zeta; l_s, p_i) \frac{x_\zeta}{\sqrt{x_{\zeta}^2 + x_{\xi}^2}} dx_\zeta
    \label{eq:Proj}
\end{equation}
where $x_\zeta = \zeta/l_s$, $x_\xi = \xi/l_p$, and $p_i$ are the parameters describing the intrinsic three-dimensional (elliptical) radial profile. This projects the three-dimensional ellipsoidal distribution of the ICM thermodynamic quantities and the gravitational potential ($F_{3D}$) onto the sky plane to produce model two-dimensional maps ($F_{2D}$) of the observables. We perform this integration numerically with logarithmic spacing of $x_\zeta$ to achieve finer sampling in the inner radial regions for better precision. The three-dimensional profiles for the physical quantities corresponding to the direct observables will be described in the following sections. 

\subsection{ICM Modeling}
\subsubsection{Electron Density and Pressure Profiles}  \label{sec:ne&Pe}
We assume the same electron density and pressure profiles as in K24. The electron density profile is defined as 
\begin{equation}
    n_e(\zeta) = n_0 \left(\frac{\zeta}{\zeta_c}\right)^{-\eta_e} \left[1 + \left(\frac{\zeta}{\zeta_c}\right)^2\right]^{-3 \beta_e / 2 + \eta_e / 2} \left[1 + \left(\frac{\zeta}{\zeta_r}\right)^3 \right]^{-\gamma_e/ 3}, 
    \label{eq:electron_density}
\end{equation}
where $n_0$ is the central electron density, $\zeta_c$ is the core radius, and $\zeta_t$ is the tidal radius ($\zeta_t > \zeta_c$). The parameters ($\beta_e, \eta_e, \gamma_e$) represent the power law exponent of the electron density distribution for the intermediate, inner, and external slope of the profile, respectively \citep{Vikhlinin2006, Ettori_2009}.

We use a generalized Navarro-Frenk-White (gNFW) profile \citep{Navarro1997, Nagai2007, Arnaud2010} to model the electron pressure profile. This is given by
\begin{equation}
    P_e(\zeta) = \frac{P_0}{(c_{500} \frac{\zeta}{R_{500c}})^{\gamma_p} [1 + (c_{500}\frac{\zeta}{R_{500c}})^{\alpha_p}]^{(\beta_p - \gamma_p)/\alpha_p}}.
    \label{eq:electron_pressure}
\end{equation}
Similar to the electron density, the parameters ($\gamma_p, \alpha_p, \beta_p$) represent the power law exponent for the central ($r \ll r_s$), intermediate ($r \sim r_s = R_{500c}/c_{500c}$), and outer ($r \gg r_s$) regions.

\citet{Nagai2007} note that that there are strong degeneracies between the pressure profile parameters, which generally prevent meaningful constraints when all are varied, so we fix the value $c_{500}$ to 1.4 and $\gamma_p$ to 0.3, following \citet{Sayers2023}. Additionally, for the demonstration in this paper, described in Section~\ref{sec:data}, the X-ray and SZ data are only fitted within the value of $R_{500c}$ derived by \planck, equal to 7.4\arcmin\ \citep[1.37~Mpc,][]{Planck2016_SZ2}. The parameter $\beta_p$ primarily influences the shape of the pressure profile beyond this radius, and it is thus not well constrained by this analysis. We therefore use the parameterization for $\beta_p$ from \citet{Sayers2023}, and so consequently our model has only two free parameters, P$_0$ and $\alpha_p$. To better assess if the resulting parametrization of the pressure profile adequately describes the observed data, we performed a series of tests to determine if an additional third free parameter is justified based on the \textit{F}-statistic. In brief, we find that none of the considered permutations of three free parameters in Eq.~\ref{eq:electron_pressure} result in a statistically significant improvement in fit quality compared to the baseline fit with two free parameters, quantified by $p$-values $\ge 0.2$ in all cases.

\begin{table*}[!t]
\caption{\label{tab:params} Model Parameters and Assumed Priors}
    \centering
    \begin{tabular}{ccp{0.5\textwidth}l}
    \hline\hline
    Parameter & Units & Description & Default Prior \\
    \hline
    \multicolumn{4}{c}{Geometrical Parameters of a Triaxial Ellipsoid} \\
    \hline
    $q_{\text{\tiny ICM,1}}$ &  & Minor-to-major axial ratio of the ICM distribution & $\mathcal{U} (0, 1)$ \\
    $q_{\text{\tiny ICM,2}}$ &  & Intermediate-to-major axial ratio of the ICM distribution & $\mathcal{U} (q_{\text{\tiny ICM,1}}, 1)$ \\
    $\cos \theta$ &  & Cosine of the inclination angle of the ellipsoid major axis & $\mathcal{U} (0, 1)$ \\
    $\varphi$ & deg & Second Euler angle & $\mathcal{U}$ (-$\pi/2$, $\pi/2$) \\
    $\psi$ & deg & Third Euler angle & $\mathcal{U}$ (-$\pi/2$, $\pi/2$) \\ \hline
    \multicolumn{4}{c}{Electron Density Profile (Equation~\ref{eq:electron_density})
    } \\ \hline
    $n_0$ & cm$^{-3}$ & Central scale density of the distribution of electrons & $\mathcal{U} (10^{-6}, 10)$ \\
    $\zeta_c$ & kpc & Ellipsoidal core radius of the gas distribution & $\mathcal{U} (0, 10^{3})$ \\
    $\zeta_t$ & Mpc & Ellipsoidal truncation radius of the gas distribution ($\zeta_t > \zeta_c$) & $\mathcal{U}$ ($\zeta_c/10^3$, 3) \\
    $\beta_e$ &  & Slope of the gas distribution (in the intermediate region) & $\mathcal{U} (0, 3)$ \\
    $\eta_e$ &  & Slope of the gas distribution (inner) & $\mathcal{U} (0, 1)$ \\
    $\gamma_e$ &  & Slope of the gas distribution (outer) & $\mathcal{U} (0, 5)$ \\ \hline
    \multicolumn{4}{c}{Gas Pressure Profile (Equation~\ref{eq:electron_pressure})
    }\\ \hline
    P$_0$ &  & Normalization for the gNFW pressure profile & $\mathcal{U} (0, 10^2)$ \\
    $c_{500}$ &  & Pressure profile concentration ($r \sim r_{\text{\tiny s}} = R_{500c} / c_{500}$) & $\delta(1.4)$ \\
    $\gamma_p$ &  & Slope parameter for central region ($r \ll r_{\text{\tiny s}}$)  & $\delta(0.3)$ \\
    $\alpha_p$ &  & Slope parameter for intermediate region ($r \sim r_{\text{\tiny s}}$) & $\mathcal{U} (0, 5)$ \\
    $\beta_p$ &  & Slope parameter for outer region ($r \gg r_{\text{\tiny s}}$) & $\delta(5.4)$ \\ \hline
    \multicolumn{4}{c}{Total Matter Density Profile} \\ \hline
    \mtwoc & 10$^{14}$\msun & Total mass  & $\mathcal{U} (1, 100)$ \\
    \ctwoc &  & Matter density profile concentration & $\mathcal{N}(\ln(c_{200c}^{\star}), 0.16)$ \\ [2px] \hline
    \hline
    \end{tabular}
    \tablefoot{For concentration, we assume a log-normal prior centered on the mass-dependent value of \ctwoc\ obtained from \citet{Diemer2019}, which we denote as \ctwoc$^{\star}$. The width of the prior, equal to 0.16, is also based on the analysis of \citet{Diemer2019}.}
    \vspace{2pt}
\end{table*}
\subsubsection{Sunyaev-Zel'dovich effect and X-ray observables } \label{sec:SZX_obs}
In this section we summarize the manner in which the assumed electron density and pressure profiles from the previous section relate to the gas observables. The SZ effect is characterized by the Compton-\textit{y} parameter, which is proportional to the integrated electron pressure along the line of sight, 
\begin{equation}
    y \equiv \frac{\sigma_T}{m_e c^2} \int_\parallel P_e dl = \frac{\sigma_T k_B}{m_e c^2} \int_\parallel n_e T_e dl,
    \label{eq:compton_y}
\end{equation}
where $\sigma_T$ is the Thomson cross-section and $k_B$ is the Boltzmann constant.

X-ray observations are primarily sensitive to the surface brightness of the ICM due to thermal Bremsstrahlung,
\begin{equation}
    \text{SB} = \frac{1}{4 \pi (1+z)^3} \int_\parallel n^2_e \Lambda(T_e, Z) dl.
    \label{eq:SB}
\end{equation}
A grid of values for $\Lambda(T_e, Z)$ is precalculated using using the Python package \verb|pyproffit| \citep{Eckert2020}, and this table is then interpolated over $T_e$ and $Z$ when evaluating the model. The emissivity calculation takes into account the instrument response within the energy band [0.7-1.2] keV and the primordial helium fraction (He/H), equal to $n_H = 7.72\times10^{-2}$ for Abell 1689 \citep{Bartalucci2023}.  

Projected maps of ICM temperature are obtained via spectroscopic fits to the X-ray data. The spectroscopic temperature is approximated within our model based on the formalism of \citet{Mazzotta2004} as 
\begin{equation}
    T_{\text{sp}} = \frac{\int W T_e dV}{\int W dV} \text{ keV}; W = \frac{n_e^2}{T_e^{3/4}},
\end{equation}
which is valid for $T_e \geq$ 3 keV.

\subsection{Mass Reconstruction} \label{sec:mass_reconstruction}
\subsubsection{Gravitational Potential Profile}
Rather than assuming the total density profile is elliptically-symmetric within our triaxial basis, we assume that the gravitational potential follows such a geometry. This decision is motivated in part by simulations, which suggest it is better to fit the potential because it is rounder and smoother than the total density \citep{Stapelburg2022}. In addition, it also allows us to make the assumption that the isopotential surfaces follow the ICM isodensity surfaces, which is strictly true in the case of an idealized pressure distribution \citep{Fox2002} and found to be approximately true in numerical simulations of galaxy clusters \citep{Lau2011}. Thus, the triaxial ellipse describing the gravitational potential has an identical shape to the ICM ($q_{\text{pot}} = q_{\text{ICM}}$), including co-alignment of the principal axes. 

The gravitational potential can be determined from the total density by inverting Poisson's equation: 
\begin{equation}
    \Phi = 4 \pi G \Delta^{-1} \rho,
    \label{eq:grav_pot_sph}
\end{equation}
where $\rho$ is the total density. This calculation is performed numerically, so any functional form for $\rho$ is allowed. Because we assume that the isopotential surfaces are elliptically symmetric with axial ratios that are constant with radius, surfaces of constant density will in general have axial ratios that vary with radius. Thus, it is not possible to describe $\rho$ using an analytic form within this elliptical basis. So, we initially invert Eq.~\ref{eq:grav_pot_sph} assuming a spherically-symmetric $\rho(r)$ based on:
\begin{equation}
    \Phi(r) = 4 \pi G \Delta^{-1} \rho(r) = 4\pi G \int \frac{1}{r^{\prime\prime2}}dr^{\prime\prime} \int dr^{\prime} r^{\prime} \rho(r^{\prime}).
\end{equation} 
We then evaluate this numerically tabulated function for $\Phi(r)$ on a three-dimensional elliptical grid, with the radial coordinate being $\zeta$. This ensures that we obtain an elliptical gravitational potential $\Phi(\zeta)$ that has identical axial ratios to those of the ICM. For this work, we assume an the initial spherically-symmetric $\rho(r)$ follows the NFW form with
\begin{equation}
    \rho(r) = \frac{\tilde{\rho}_s}{(r/\tilde{r}_s)(1 + r/\tilde{r}_s)},
    \label{eq:NFW_sph}
\end{equation}
where $\tilde{\rho}_s$ and $\tilde{r}_s$ are the characteristic density and scale radius. We emphasize that $\Phi(\zeta)$ is the quantity that is projected along the line of sight, as described in Section~\ref{sec:Projection}, to obtain our WL observables. 

It is common to quantify the total density distribution by \mtwoc, which is the total mass within a sphere of radius \rtwoc, and a concentration \ctwoc~$= \text{R}_{200c} / r_s$ where $r_s$ denotes the radius where the logarithmic slope of the density profile is equal to $-2$. To obtain these quantities from our elliptical potential $\Phi(\zeta)$ we proceed as follows. The underlying density distribution in three-dimensions corresponding to $\Phi(\zeta)$ is found by solving Poisson's equation numerically in Cartesian coordinates as
\begin{equation}
    \rho(\zeta(x,y,z)) = \frac{1}{4\pi G} \Delta_{xyz} \Phi(\zeta(x,y,z)).
    \label{eq:NFW_ell}
\end{equation}
As noted above, the axial ratios corresponding to constant values of $\rho(\zeta)$ will in general vary with radius. To obtain \mtwoc, we spherically integrate $\rho(\zeta)$ in three-dimensions. To determine \ctwoc, we spherically average $\rho(\zeta)$ to calculate the scale radius $r_s$ as defined above. We note that $\tilde{\rho}_s$ and $\tilde{r}_s$ are distinct from the $\rho_s$ and $r_s$ associated with this spherically averaged profile of $\rho(\zeta)$, from which we find \mtwoc\ and \ctwoc. Therefore, $\tilde{\rho}_s$ and $\tilde{r}_s$ are not physically meaningful in this context, but are used here to connect \mtwoc\ and \ctwoc\ back to Eq.~\ref{eq:grav_pot_sph}. For a given set of input parameters ($q_{\text{ICM,1}}$, $q_{\text{ICM,2}}$, $\tilde{\rho}_s$ and $\tilde{r}_s$), we thus obtain unique values of the physical quantities of interest, \mtwoc\ and \ctwoc. Spanning the full set of allowed input parameter values, we then tabulate the relation between them and \mtwoc\ and \ctwoc. By interpolating this table, we can then connect any combination of \mtwoc\ and \ctwoc\ to the underlying values $\tilde{\rho}_s$ and $\tilde{r}_s$ for a given $q_{\text{ICM,1}}$ and $q_{\text{ICM,2}}$. It is important to note that, although our formalism includes an NFW $\rho(r)$ as a starting point, and we quantify the mass distribution via \mtwoc\ and \ctwoc, the underlying density distribution obtained from Eq.~\ref{eq:NFW_ell} does not strictly follow an NFW distribution.

\subsubsection{Weak lensing observables} \label{sec:WLobs}
The effect of WL by a matter distribution on a background source is characterized by convergence, $\kappa$, and shear, $\gamma$. Convergence is an isotropic rescaling of the source, while shear is a stretching of the intrinsic shape of the source along some preferred axis. The shear introduces a quadrupole anisotropy in the background images, which can be observed from ellipticities of background galaxies. In general, the observable quantity for quadrupole WL is not $\gamma$ but the reduced gravitational shear \citep{Umetsu2020rev}, defined as
\begin{equation}
    g = \frac{\gamma}{1 - \kappa}.
    \label{eq:reduced_shear}
\end{equation}
In our analysis, we use reduced shear maps to reconstruct the density profile of the cluster. The total matter density profile is related to shear through series of calculations that will be described now. An extended distribution of matter is characterized by its effective lensing potential, which is the rescaled projection of the three-dimensional Newtonian potential on the lens plane: 
\begin{equation}
    \Psi = \frac{D_{LS}}{D_L D_S} \frac{2}{c^2} \int \Phi dl,  
    \label{eq:LensingPot_Int}
\end{equation}
where $D_L$, $D_S$, and $D_{LS}$ are the angular diameter distances from the observer to the lens, which is the galaxy cluster in this case, the observer to the source, and the lens to the source, respectively. The shear is defined as a complex vector with spin-2 rotational symmetry $\vec{\gamma} = (\gamma_1, i\gamma_2)$, on the lens plane, and its components are calculated from the lensing potential as
\begin{equation}
    \gamma_1 = \frac{1}{2}(\Psi_{11} - \Psi_{22}) \quad\mathrm{and}\quad \gamma_2 = \Psi_{12} = \Psi_{21},
    \label{eq:shear}
\end{equation}
using the shorthand notation $ \frac{\partial^2 \Psi(\vec{x})}{\partial x_i \partial x_j} \equiv \Psi_{ij}$. The convergence is
\begin{equation}
    \kappa = \frac{1}{2} \nabla^2 \Psi.
    \label{eq:kappa}
\end{equation}
The two components of the reduced shear can then be constructed from Eqs.~\ref{eq:shear} \& \ref{eq:kappa}, as defined in Eq.~\ref{eq:reduced_shear}. 

While our fit is based on the two-dimensional projected two-component shear ($\gamma_1$ and $\gamma_2$), we also construct one-dimensional radial profiles of the tangential shear to better enable visual comparisons between the observed data and model (see Section~\ref{sec:Results}). We briefly summarize here our calculation of the tangential shear. For more details, see \citet{Umetsu2020rev}. In a polar coordinate basis $(\theta,\phi)$ centered on the lens, the shear signal can be deconstructed into two components, the tangential shear $\gamma_+$, and the $45\degree$ rotated cross shear $\gamma_\times$: 
\begin{equation}
\begin{split}
    \gamma_+(\theta,\phi) = -\gamma_1(\theta)\cos2\phi - \gamma_2(\theta)\sin2\phi, \\ 
    \gamma_\times(\theta,\phi) = +\gamma_1(\theta)\sin2\phi - \gamma_2(\theta)\cos2\phi.
\end{split}
\end{equation}
For an arbritary choice of the coordinate center, the azimuthally averaged components of the tangential and cross shear are expressed as \citep[e.g.,][]{Umetsu2020rev} 
\begin{equation}
    \begin{split}
        \gamma_+(\theta) = \frac{\Delta \Sigma(\theta)}{\Sigma_{\text{crit}}}, \\
        \gamma_\times(\theta) = 0, 
    \end{split}
\end{equation}
where $\Sigma_{\text{crit}} = \frac{c^2}{4\pi G} \frac{D_\text{s}}{D_{\text{l}} D_{\text{ls}}}$, and we have introduced the excess surface mass density $\Delta \Sigma(\theta)$, defined as the difference between the average surface mass density $\overline{\Sigma}(\theta)$ within a circle of radius $\theta$ and the local surface mass density: 
\begin{equation}
    \begin{split}
        \Delta \Sigma(\theta) = \overline{\Sigma}(\theta) - \Sigma(\theta), \\
        \overline{\Sigma}(\theta) = \frac{2}{\theta^2} \int_0^\theta \theta' \Sigma(\theta')d\theta'.
    \end{split}
\end{equation}
The tangential reduced shear is then defined as 
\begin{equation}
    g_+(\theta) = \frac{\gamma_+}{1 - \kappa(\theta)},
    \label{eq:gt}
\end{equation}
where we calculate the convergence as $\kappa = \Sigma/\Sigma_{\text{crit}}$. Because triaxial halos have elliptical isodensity contours in projection on the sky, Eq.~\ref{eq:gt} gives a good approximation to describe the WL signal for regular clusters \citep{Umetsu2020rev}. To compare our model to the observed tangential shear data, we calculate the mean tangential shear in the non-linear regime using the formulation of \citet{Seitz+Schneider1997}: 
\begin{equation}
    \langle g_+(\theta) \rangle = \frac{\langle \gamma_+ (\theta) \rangle}{1 - f_l\langle \kappa(\theta) \rangle},
    \label{eq:gt_avg}
\end{equation}
with 
\begin{equation}
    f_l = \frac{\langle \Sigma_{\text{crit}}^{-2} \rangle}{\langle \Sigma_{\text{crit}}^{-1} \rangle ^2}.
\end{equation}
We again emphasize that our model fits two-dimensional component reduced shear maps $g_1$ and $g_2$ and that we perform this calculation of $\langle g_+ \rangle$ as a consistency check. 

\subsection{Projection of Three-Dimensional Quantities} \label{sec:Projection}

The SZ, X-ray, and WL observables (Eqs.~\ref{eq:compton_y}, \ref{eq:SB}, and \ref{eq:LensingPot_Int}) are modeled as projections of the three-dimensional profiles parameterized by the ellipsoidal radial coordinate $\zeta$ and its scaled equivalent $x_{\zeta}$. As detailed below, all of these projections are performed numerically with sufficient sampling to ensure a fractional precision of approximately $10^{-3}$ (or better), far below measurement uncertainties. In the case of the SZ and X-ray observables, the three-dimensional electron pressure and density, $F_{3D}(x_\zeta; l_s, p_i)$, can be written analytically and calculated following Eq.~\ref{eq:Proj}. The model Compton-\textit{y} parameter is
\begin{equation}
    y_{\text{model}}(x_\xi;l_p,p_i) = (2l_p e_\parallel)\left( \frac{\sigma_T}{m_e c^2} \right) \int_{x_\xi}^\infty P_e(x_\zeta) \frac{x_\zeta}{\sqrt{x_\zeta^2 - x_\xi^2}} dx_\zeta.
\end{equation}
Similarly, the X-ray SB model is 
\begin{multline}
    \text{SB}_{\text{model}}(x_\xi;l_p,p_i) = (2l_p e_\parallel) \frac{1}{4 \pi (1+z)^3} \\
    \times \int_{x_\xi}^{\infty} n_e^2(x_\xi) \Lambda\left( T_e(x_\zeta), Z(x_\zeta) \right) \frac{x_\zeta}{\sqrt{x_\zeta^2 - x_\xi^2}} dx_\zeta, 
\end{multline}
where the electron temperature for the cooling function is calculated as
\begin{equation}
    T_e(x_\zeta) = \frac{P_e(x_\zeta)}{n_e(x_\zeta) k_B},
\end{equation}
and we use a constant metallicity of 0.3 of the solar abundance.

The lensing potential is modeled as a projection of the three-dimensional Newtonian potential. In this case, the three-dimensional function projected to make the two-dimensional maps in Eq.~\ref{eq:Proj} is a numerically calculated function. The model lensing potential is then

\begin{multline}
        \Psi_{\text{model}}(x_\xi; l_p, p_i) = \\
        (2l_p e_\parallel) \left(\frac{D_{LS}}{D_L D_S}\right) \int_{x_\xi}^\infty \Phi(x_\zeta; l_s, p_i) \frac{x_\zeta}{\sqrt{x_{\zeta}^2 - x_{\xi}^2}} dx_{\zeta}.
        \label{eq:model_LP}
\end{multline}
Partial derivatives are taken of this model lensing potential according to Eq.~\ref{eq:shear} to produce the model shear maps. 

The projection integration for each of the observables can be computationally expensive, so to expedite the calculation, we created a linearly spaced sample of the (normalized) elliptical radius $x_\xi$ and interpolated the integration results while generating a model. Additionally, given the centrally-peaked profiles for the SZ, X-ray SB, and WL shear, the integration was updated to be performed in log-space since it provides the best combination of accuracy and run-time. Further details on this change can be found in Appendix~\ref{sec:appen:K24_compare}. 

For our application to Abell 1689 in this analysis, the observational data include two-dimensional images of the SZ, X-ray SB, and WL shear signals, as well as one dimensional projected radial profiles of X-ray SB and temperature data. The triaxial model detailed in this section generates analogous images and radial profiles based on the model parameters in Table~\ref{tab:params}. The observed and model-generated data can be directly compared within our fitting pipeline to constrain the parameter values, and the method employed for doing this will be discussed in the following section. 

\subsection{Fitting formalism}
The $\chi^2$ statistic is used to define the likelihood of the model. We use \verb|emcee| \citep{emcee}, a Python-based affine-invariant ensemble Markov chain Monte Carlo \citep[MCMC;][]{Goodman2010} package, for the model fitting process. The posterior distribution of the parameter values describing the triaxial model was constructed through MCMC sampling \citep{Hogg2018}.  

We use the same definitions of $\chi^2$ for the SZ, X-ray SB and X-ray temperature data as in K24. For the SZ map, this is 
\begin{equation}
    \chi^2_{\text{SZ}} = \sum_{i,j=1}^{N_y} \left[y_i - \hat{y}_i \right]\left(C_{\text{SZ}}^{-1} \right)_{i,j} \left[y_j - \hat{y}_j \right],
\end{equation}
where $\hat{y}_i$ is the model Compton-\textit{y} within a pixel, $y_i$ is the observed value, and $C_{\text{SZ}}$ is the covariance matrix. For the X-ray temperature, the $\chi^2$ function is similarly defined as 
\begin{equation}
    \chi^2_\text{T} = \sum_{i=1}^{N_T} \left( \frac{T_{\text{sp},i} - \hat{T}_{\text{sp},i}}{\delta T_{\text{sp},i}} \right)^2, 
\end{equation}
where $\hat{T}_{\text{sp},i}$ is the model spectroscopic temperature within a temperature bin in a radial profile, and $T_{\text{sp},i}$ is the observed value with uncertainty $\delta T_{\text{sp},i}$. 

A dual approach employing two-dimensional and one-dimensional fits for the X-ray SB is used. We utilize two-dimensional data within the circular region that encloses 80~per cent of the emission and a one-dimensional projected profile for the radial region outside of this, where the background and the source emission is comparable and signal-to-noise ratio is relatively low. A one-dimensional analysis in the exterior region is further motivated by a desire to mitigate biases in measuring X-ray SB caused by gas clumping, which can be achieved by taking the azimuthal median, rather than mean, of the observed data as suggested by \citet{Eckert2015}. To generate one-dimensional X-ray SB profiles from the model, an azimuthal mean in annular bins is obtained from the two-dimensional projected model image in the exterior region. For the model, a mean averaging is approximately equivalent to median averaging, since the model is smooth and does not contain any clumps. The combined likelihood from the two radial regions becomes 
\begin{equation}
    \chi^2_\text{SB} = \chi^2_\text{SB,1D} + \chi^2_\text{SB,2D}
\end{equation}
where
\begin{equation}
    \chi^2_{\text{SB,1D}} = \sum_{i=1}^{N_\text{SB,1D}} \left( \frac{S_{\text{X,1D,}i} - \hat{S}_{\text{X,1D,}i}}{\delta S_{\text{X,1D,}i}} \right)^2
\end{equation}
and 
\begin{equation}
    \chi^2_{\text{SB,2D}} = \sum_{i=1}^{N_\text{SB,2D}} \left( \frac{S_{\text{X,2D,}i} - \hat{S}_{\text{X,2D,}i}}{\delta S_{\text{X,2D,}i}} \right)^2.
\end{equation}
We note that there is no overlap between the regions used for the one-dimensional and two-dimensional analyses. Here $\hat{S}_{\text{X},i}$ is the model SB, and $S_{\text{X,}i}$ and $\delta S_{\text{X,}i}$ are obtained from the observational data under the assumption that the uncertainties are Gaussian, which should be approximately valid given the large number of counts within each two-dimensional pixel and one-dimensional radial bin. 

The $\chi^2$ function for the two-dimensional reduced shear is 
\begin{equation}
    \chi_{\text{WL}}^2 = \sum_{i,j=1}^{N_g} \sum_{\alpha,\beta=1}^2 \left[g_{\alpha,i} - \hat{g}_{\alpha,i}\right] \left(C_{\text{WL}}^{-1}\right)_{\alpha\beta,ij} \left[g_{\beta,j} - \hat{g}_{\beta,j}\right]
    \label{chi2_WL}
\end{equation}
where $\hat{g}_i$ is the model reduced shear map within a pixel, $g_i$ is the observed value, and $C_{\text{WL}}$ is the covariance matrix. In our error analysis we account for statistical noise (shape noise) and cosmic variance in the noise covariance matrix as $C_{\text{WL}} = C^{\text{stat}} + C^{\text{lss}}$. The shape noise is calculated according to Eq.~\ref{eq:bin_shearvar} and $C^{\text{lss}}$ is the noise covariance due to uncorrelated large scale structure (LSS) along the line of sight \citep{Hoekstra2003}. The contribution from $C^{\text{stat}}$ dominates over $C^{\text{lss}}$, with the LSS noise amplitude being $\sim$1/100 of the shape noise amplitude, however, the impact of cosmic noise is most important to include when the cluster signal itself is small \citep{Hoekstra2003}. This is the case for the large cluster radii we are interested in probing. The components of the $C^{\text{lss}}$ matrix are calculated following the prescription outlined in \citet{Kaiser1992}, \citet{Hu_White2001}, and \citet{Schneider2002}. We use a lensing power spectrum $C^{\epsilon\epsilon}(l)$ calculated for a source population at $z_s$ = 1, using the cosmological parameters from the Wilkinson Microwave Anisotropy Probe (WMAP) Nine-Year results \citep{WMAP9}.  

Using all three observational probes, the total $\chi^2$ statistic becomes 
\begin{equation}
    \chi^2_{\text{X+SZ+WL}} = \chi^2_{\text{SZ}} + \chi^2_{\text{T}} + \chi^2_{\text{SB}} + \chi^2_{\text{WL}}
\end{equation}
and the MCMC is used to sample $\chi^2_{\text{X+SZ+WL}}$ within the parameters space near the best fit. While the total $\chi^2$ from all of the observables is evaluated in our fit, to good approximation the three ICM observables constrain the parameters associated with the three-dimensional geometry. Given this geometry, the parameters of the total matter density are constrained entirely by the WL data and the parameters of the ICM density are primarily constrained by the X-ray SB. The combination of the SZ and X-ray temperature data then constrain the parameters of the ICM pressure.

\begin{figure*}
    \centering
    \includegraphics[width=0.32\linewidth]{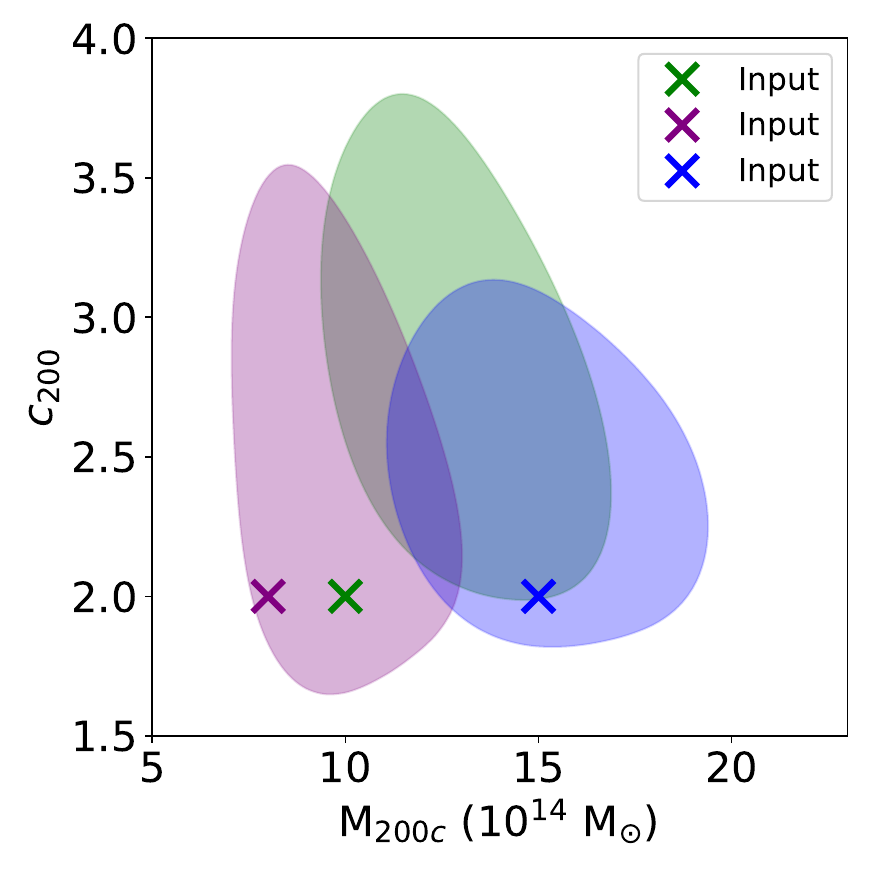}
    \includegraphics[width=0.32\linewidth]{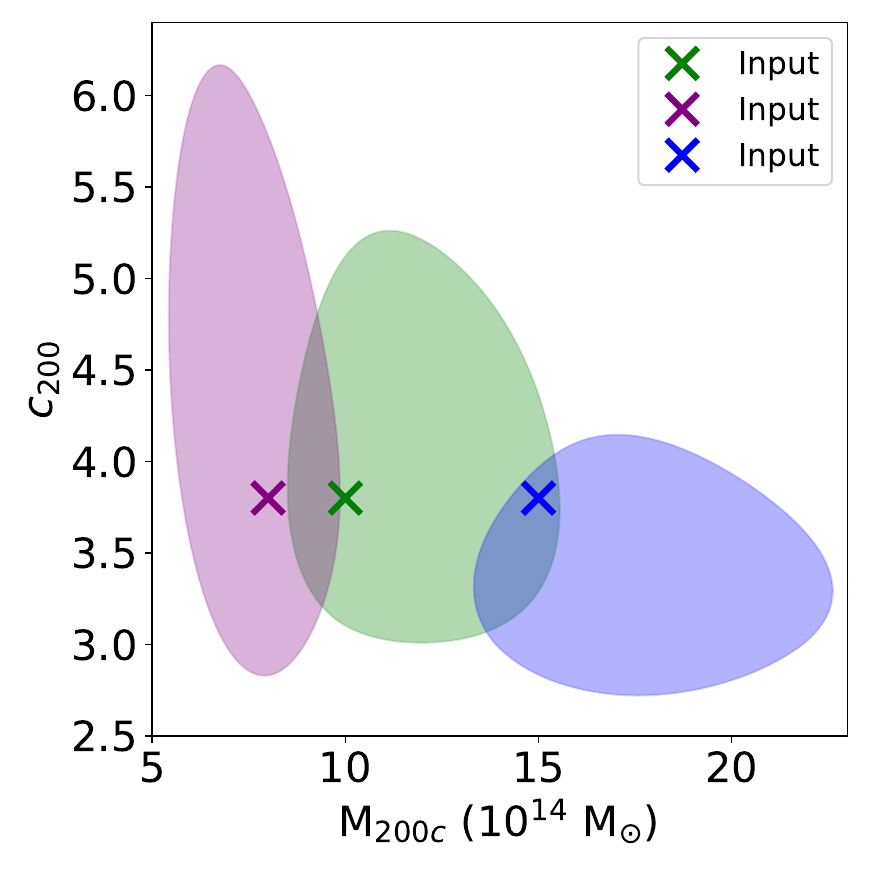}
    \includegraphics[width=0.32\linewidth]{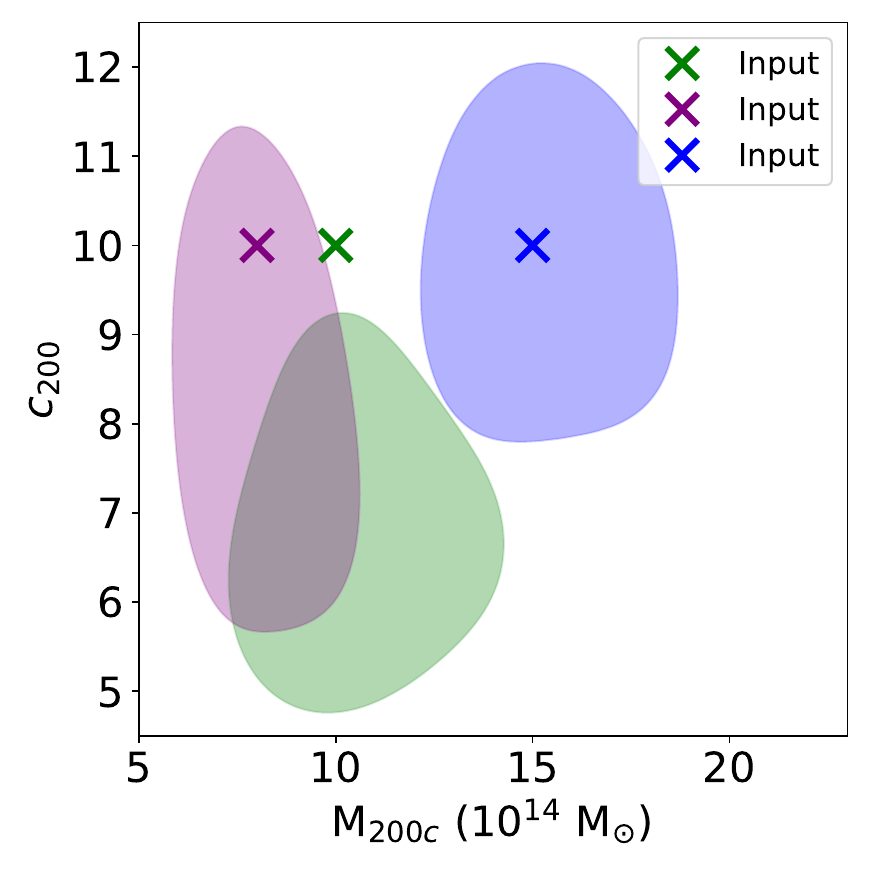}
    \caption{The posterior distribution of \mtwoc\ and \ctwoc\ obtained from fits to mock observations. The shaded region for each posterior represents the 68~per cent credible region. The corresponding input values are indicated by the $\times$ symbol. In all cases the recovered \mtwoc\ value is consistent with the input, within noise fluctuations. This is also true for \ctwoc\ when the input value is \ctwoc$=3.8$, as this value is close to the center of the log-normal prior utilized. For the other two cases, there is a small, but measurable bias in the recovered \ctwoc\ values.}
    \label{fig:toymodel_cM_results}
\end{figure*}

\subsubsection{Priors} \label{sec:priors}
For all the free model parameters except \ctwoc, we apply uniform priors that are uninformative other than with regard to physical boundaries (e.g., $q_{ICM,1}$ must take values between 0--1, see Table~\ref{tab:params}). It has been demonstrated that with WL alone, it is very difficult to simultaneously constrain both \mtwoc\ and \ctwoc\ \citep[e.g.,][]{Meneghetti2010_SL, Postman2012}, and so it is common to place an informative prior on \ctwoc, generally in the form of a $\delta$-function or based on $c$-$M$ relations obtained from simulations \citep[e.g.,][]{Applegate2014, Giocoli2024, Hoekstra2015}. We adopt the latter approach, employing a log-normal prior on \ctwoc, with central value set by the average $c$-$M$ relation given by the model of \citet{Diemer2019} and an assumed width based on the intrinsic scatter of this relation, corresponding to $\log(c_{200c}) \pm 0.16$. This prior captures the expected intrinsic cluster-to-cluster scatter in $c$-$M$ values in the Universe. To compute the central values of the $c$-$M$ relation, we utilized the Python package \verb|colossus| \citep{COLOSSUS}. For the span of the uniform prior we use for \mtwoc, this $c$-$M$ relation returns values for \ctwoc\ between $\sim4-6.5$. This matches simulated concentrations from both scale-free models and $\Lambda$CDM to 5~per cent accuracy \citep{Diemer2019}. 


\section{Non-thermal Pressure Calculation}  \label{sec:Pnt}
The addition of a WL analysis enables indirect measurement of the motions outside of thermal equilibrium that contribute to the total pressure of the cluster. From the constraints on the total mass density and the gas density profiles we can calculate the total pressure needed to offset gravitational collapse, \ptot, and the thermal component of pressure, \pth, in a three-dimensional triaxial basis. Because clusters are not in a state of perfect HSE, \ptot\ is in general greater than \pth\ and so the non-thermal component of pressure can be measured as the difference between \ptot\ and \pth\ under the assumption that the system is fully virialized. Within this formalism we are not sensitive to the underlying source of the non-thermal pressure support, which could be due to the combination of bulk motions, turbulence, magnetic fields, cosmic rays, or any other relevant terms. Assuming a state of generalized HSE, the total pressure is given by 
\begin{equation}
    \nabla P_{\text{tot}} = - \rho_{gas} \nabla \Phi,
    \label{eq:gHSE}
\end{equation}
where \ptot = \pth + \pnt\ and $\rho_{gas}$ is the gas density. Following the formalism of \citet{Fox2002}, for an elliptically symmetric system this reduces to 
\begin{equation}
    \frac{dP_{\text{tot}}}{d\zeta} = -\rho_{gas} \frac{d\Phi}{d\zeta}
    \label{eq:gHSE_ellip}
\end{equation}
and \ptot\ can be found by numerically integrating over the elliptical radius. We calculate $\rho_{gas}$ from our assumed electron density profile (Eq.~\ref{eq:electron_density}) through the relation
\begin{equation}
        \rho_{gas} = \mu_e m_p n_e,
        \label{eq:gas_density}
    \end{equation}
where $\mu_e$ is mean molecular weight per free electron, taken to be equal to 1.14 as in \citet{Nagai2007}, and $m_p$ is the mass of the proton. The Newtonian gravitational potential on a triaxial ellipsoid is calculated numerically as described in Section~\ref{sec:mass_reconstruction}. The thermal pressure is calculated from the electron pressure through the relation 
\begin{equation}
    P_{th} = P_e \frac{\mu_e}{\mu},
    \label{Pth}
\end{equation}
where $\mu$ is mean molecular weight, taken to be equal to 0.59 as in \citet{Nagai2007}. All of the above calculations are performed within a triaxial basis, and so we obtain a spherically-averaged radial profile of \pnt\ by azimuthally averaging \ptot\ and \pth\ in three dimensions prior to computing their difference. 

\begin{table*}
\caption{\label{tab:mock_results} Recovered Parameters from Fits to Mock Data}
    \centering
    \begin{tabular}{cc|cc|cc}
        \hline \hline \\[-8px]
        \multicolumn{2}{c}{\mtwoc\ (10$^{14}$~\msun)} & \multicolumn{2}{c}{\ctwoc} & \multicolumn{2}{c}{\pnt/\ptot\ at R=1 Mpc}\\
        Input & Recovered & Input & Recovered & Input & Recovered\\ \hline \\[-5px]
         8 & $7.43_{-1.28}^{+1.50}$ & 3.8 &  $4.33_{-0.94}^{+1.23}$ & 0.25 & $0.21_{-0.11}^{+0.11}$\\[5px]
        10 & $11.81_{-2.24}^{+2.31}$ & 3.8 &  $4.00_{-0.66}^{+0.78}$ & 0.36 & $0.43_{-0.09}^{+0.08}$\\[5px]
        15 & $17.67_{-2.88}^{+2.99}$ & 3.8 &  $3.37_{-0.41}^{+0.49}$ & 0.51 & $0.55_{-0.06}^{+0.05}$ \\[2px] \hline \\[-5px]
         8 & $8.06_{-1.44}^{+1.52}$ & 10 &  $8.23_{-1.66}^{+2.03}$ & 0.31 & $0.30_{-0.12}^{+0.14}$\\[5px]
        10 & $10.39_{-1.86}^{+2.36}$ & 10 &  $6.70_{-1.09}^{+1.62}$ & 0.42 & $0.40_{-0.10}^{+0.10}$ \\[5px]
        15 & $15.36_{-2.12}^{2.18}$ & 10 &  $9.77_{-1.09}^{+1.62}$ & 0.58 & $0.59_{-0.05}^{+0.05}$\\[2px] \hline \\[-5px]
         8 & $9.51_{-1.61}^{+2.14}$ & 2 &  $2.48_{-0.58}^{+0.64}$ & 0.40 & $0.39_{-0.12}^{+0.14}$\\[5px]
        10 & $12.69_{-2.17}^{+2.60}$ & 2 &  $2.79_{-0.51}^{+0.63}$ & 0.46 & $0.47_{-0.08}^{+0.08}$ \\[5px]
        15 & $14.81_{-2.33}^{+3.11}$ & 2 &  $2.43_{-0.40}^{+0.45}$ & 0.56 & $0.53_{-0.08}^{+0.07}$\\[2px] 
        \hline \hline 
    \end{tabular}
    \label{tab:my_label}
\end{table*}

Because the above calculation is computationally expensive, it is not feasible to perform it within the fit at every step in the MCMC. So, the calculation is evaluated outside the fit, using the set of accepted samples from the MCMC. Because of this, it is not possible to apply a prior on \pnt\ to ensure its value remains in the physically allowed region $\in [0,$~\ptot]. For this analysis, we do not obtain values outside of this range, and so this lack of a prior does not present a problem for our fits. However, we note that techniques have been established for this exact scenario, as described in \citet{Sayers2021}.
\section{Demonstration With Mock Data} \label{sec:MockData}
To validate the accuracy of our model fitting algorithm, we performed a full analysis using mock observations of galaxy clusters generated from our model using known input parameter values. A validation of the geometric and gas parameters was performed in K24, and so here we focus on testing the pipeline on mock data generated from various combinations of \mtwoc\ and \ctwoc\ values, listed in Table~\ref{tab:mock_results}, with the following identical set of geometric and ICM gas parameters for all mock clusters: ($q_{\text{ICM},1}$, $q_{\text{ICM},2}$, cos$\theta$, $\varphi$, $\psi$) = (0.6, 0.75, 0.6, -25, 60), ($n_0$, $\zeta_c$, $\zeta_t$, $\beta_e$, $\eta_e$, $\gamma_e$) = (0.01, 200, 2.5, 0.65, 0.60, 4.2), (P$_0$, $\alpha_p$) = (30, 0.8). The corresponding elongation for this triaxial geometry is $\mathcal{R}_{LP}=1.04$ . The mock clusters also share the following characteristics, which closely mimic those of observed cluster we perform our demonstration on, Abell 1689, with $z_{\text{cluster}}$ = 0.18 and $z_{\text{source}}$ = 0.83. We then consider three values for \mtwoc, (8, 10, 15)~$\times 10^{14}$~\msun, and three values for \ctwoc, (2, 3.8, 10), for a total of nine different mock clusters. From these parameters, we generate model SZ, X-ray SB and temperature maps, and WL shear maps for each cluster. 

Instrumental effects, including the point spread function (PSF) response, were incorporated in the mock observations. Noise was included in each mock observation using the error maps for the observed data. From these, we randomly sampled Gaussian noise distributions for the SZ, X-ray SB and temperature, and WL shear maps, respectively. As expected, many of the input parameters fall within the 68~per cent confidence intervals obtained from our fits, and are otherwise contained within the 95~per cent confidence intervals. We show summary plots of the values of \mtwoc\ and \ctwoc\ obtained from all 9 fits in Figure~\ref{fig:toymodel_cM_results}. In all cases, we find that the recovered values of \mtwoc\ are consistent with the input values given the expected noise fluctuations. For \ctwoc\ this is also the case when the input model has \ctwoc$=3.8$, as this value is close to the center of the log-normal prior we utilize. However, in the other two cases, the prior on \ctwoc\ results in a small, but measurable bias in the recovered values.

\begin{figure*}
    \centering
    \includegraphics[width=0.32\linewidth]{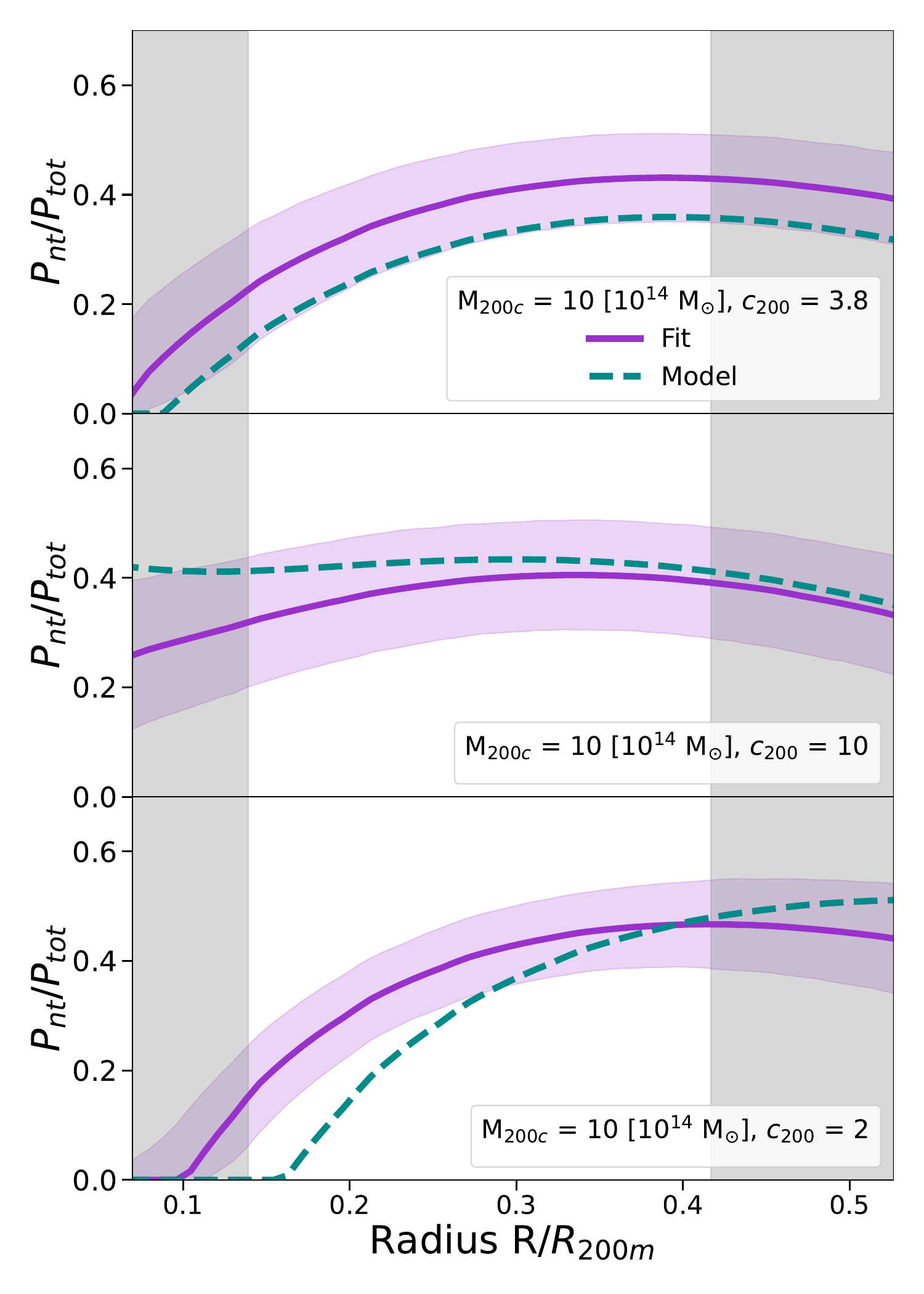}
    \includegraphics[width=0.32\linewidth]{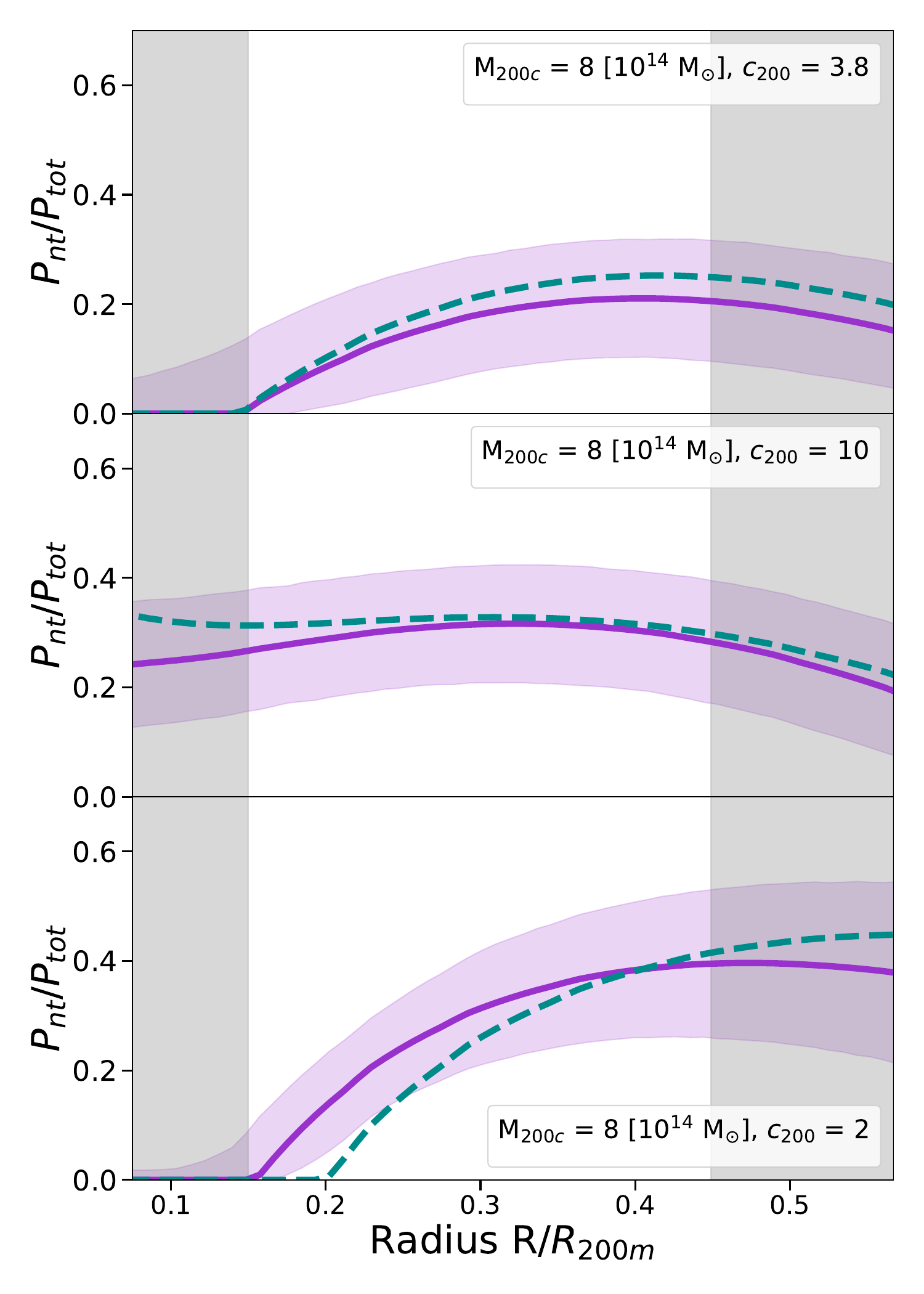}
    \includegraphics[width=0.32\linewidth]{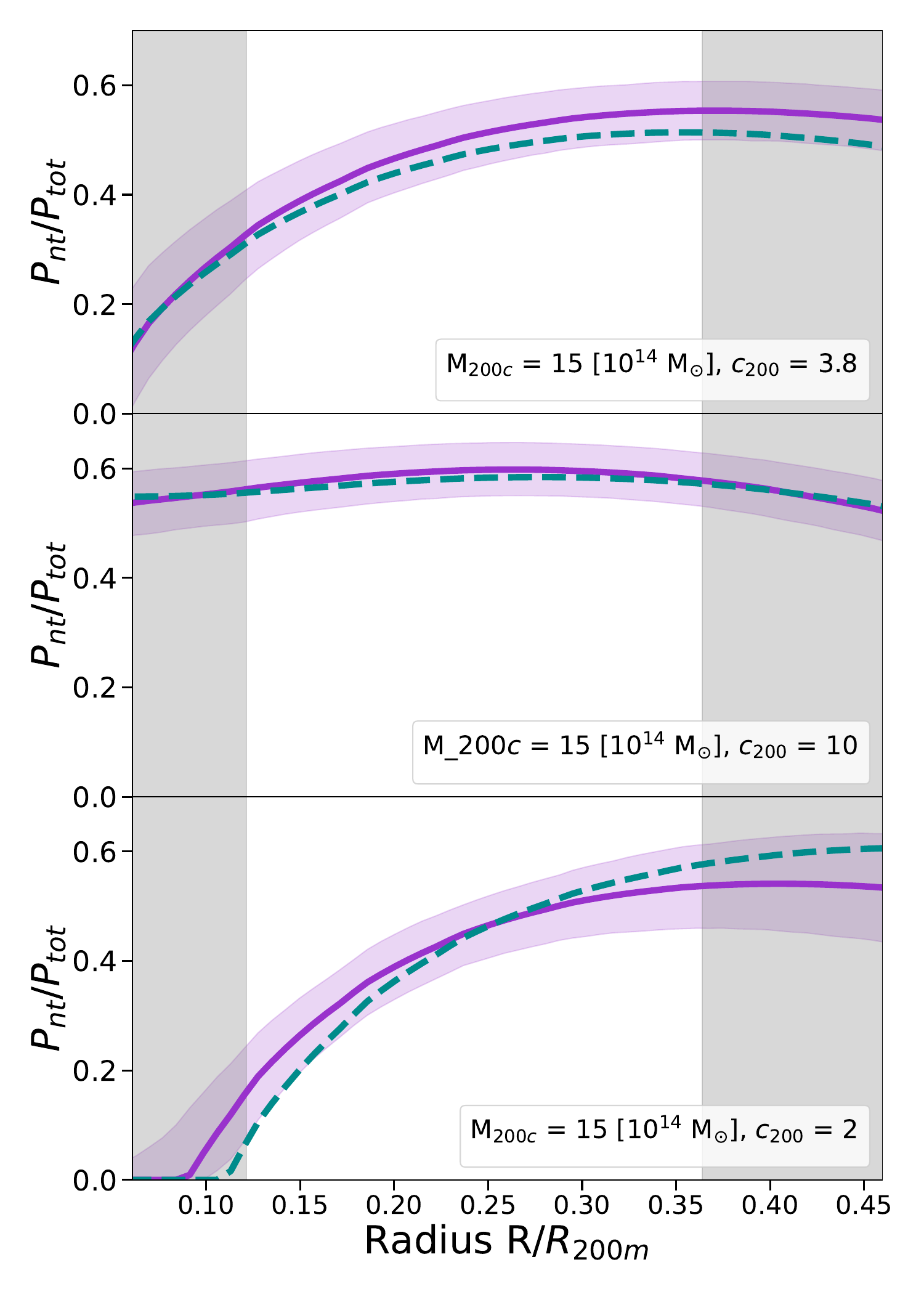}
    \caption{Radial plots of the non-thermal pressure fraction (\pnt/\ptot) recovered from fits to mock observational data, generated with a range of \mtwoc\ and \ctwoc\ values. The solid purple line represents the best fit model and the shaded purple region is the corresponding 68~per cent credible region calculated from the accepted MCMC samples. The dashed blue line is the non-thermal pressure fraction calculated from the input parameters to generate the mock data. Shaded gray denotes regions where the mass distribution may not be fully constrained from the observed WL shear signal (see Section~\ref{sec:WLdata}). We find good agreement between the recovered and input profiles for \ctwoc\ equal to 3.8 and 10 and a slight bias in the shape of \pnt/\ptot\ for \ctwoc$=2$ due to the informative prior applied to this parameter.}
    \label{fig:toymodel_Pnt}
\end{figure*}

The corresponding radial profiles of \pnt/\ptot\ were calculated using the accepted samples from the 9 fits following the approach outlined in Section~\ref{sec:Pnt}, and these radial plots are displayed in Figure~\ref{fig:toymodel_Pnt}. We find the input profile of \pnt/\ptot\ falls within the 68~per cent credible region of the recovered posterior over the full radial range for input models with \ctwoc\ equal to 3.8 and 10, and for most of the radial range for models with \ctwoc$=2$. This slight bias in the shape of the non-thermal pressure fraction for the \ctwoc$=2$ case is attributed to the the bias in the recovered \ctwoc\ values as a result of the informative prior on its value. A summary of the relevant recovered parameters from the fits to all the mock observations is given in Table~\ref{tab:mock_results}. From these, we infer that any potential bias in our fitting method is minimal compared to the measurement uncertainties. 
\section{Observational Data} \label{Data}
\label{sec:data}
In this section, we summarize the SZ, X-ray, and WL data available for our analysis, which differ from those used in K24, except for the two-dimensional X-ray SB map which is identical. These data were assembled as part of the Cluster HEritage project with \xmm: Mass Assembly and Thermodynamics at the Endpoint of Structure Formation collaboration \citep[CHEXMATE;][]{chexmate2021}. In particular for this analysis, we consider the CHEX-MATE galaxy cluster \cluster (Abell 1689), which is at a redshift of $z=0.183$. Abell 1689 is an extremely massive cluster, so it acts as a powerful gravitational lens and provides high quality lensing data. This makes it an ideal candidate for making precise mass measurements within our triaxial framework. Additionally, this cluster is well studied in the literature, allowing for detailed comparisons of these prior results with our primary measurements (mass, concentration, elongation, and non-thermal pressure content). 

\subsection{SZ and X-ray data}
\label{sec:gas_data}
\begin{figure*}
    \centering
    \includegraphics[width=0.32\linewidth]{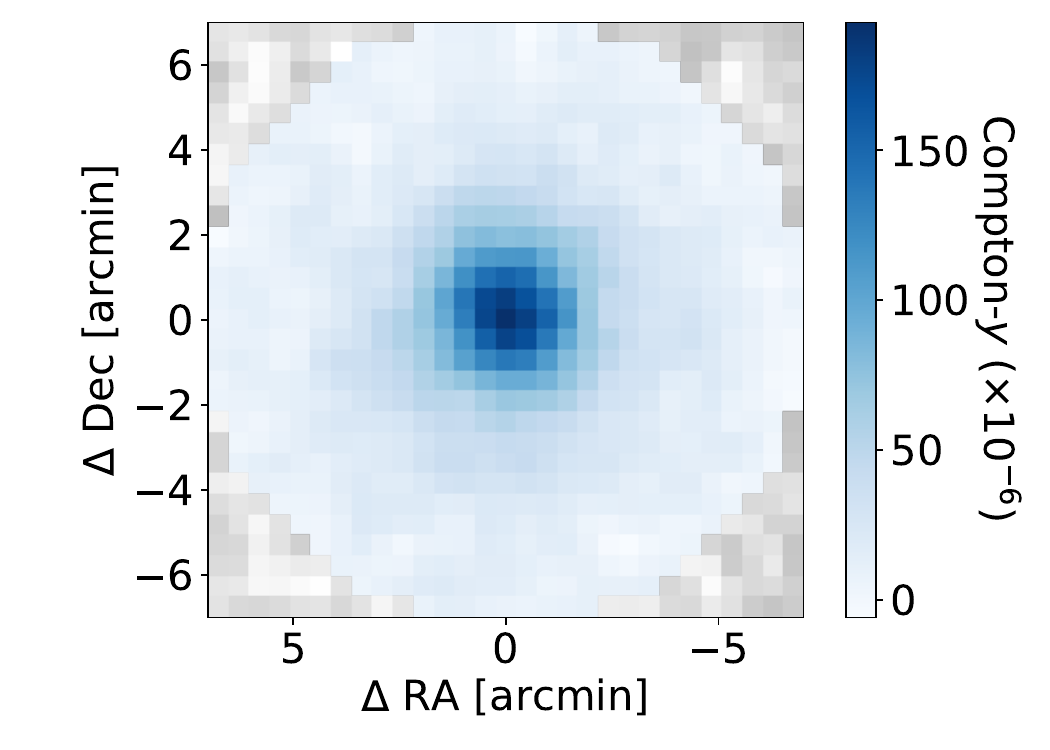}
    \includegraphics[width=0.32\linewidth]{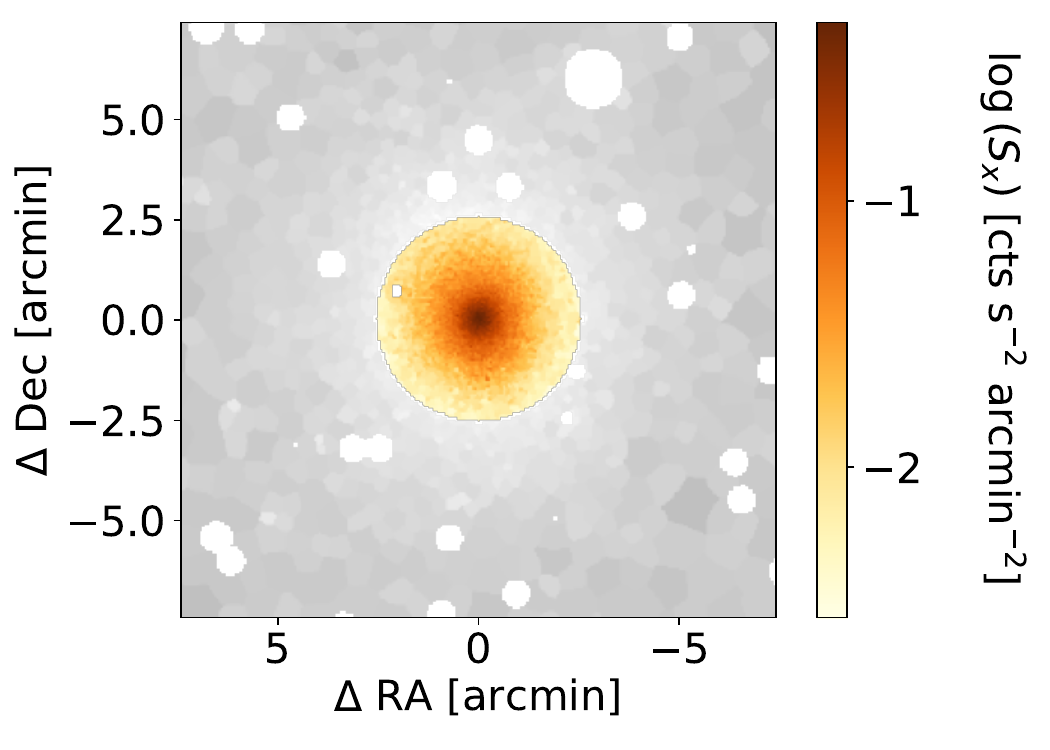}
    \includegraphics[width=0.32\linewidth]{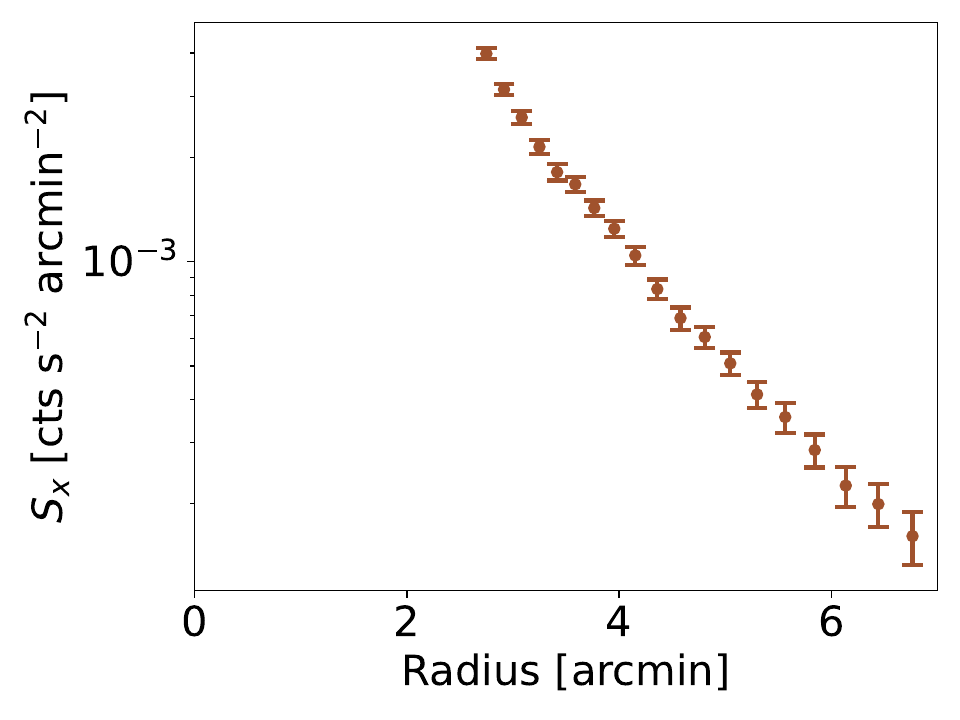}
    \\
    \includegraphics[width=0.32\linewidth]{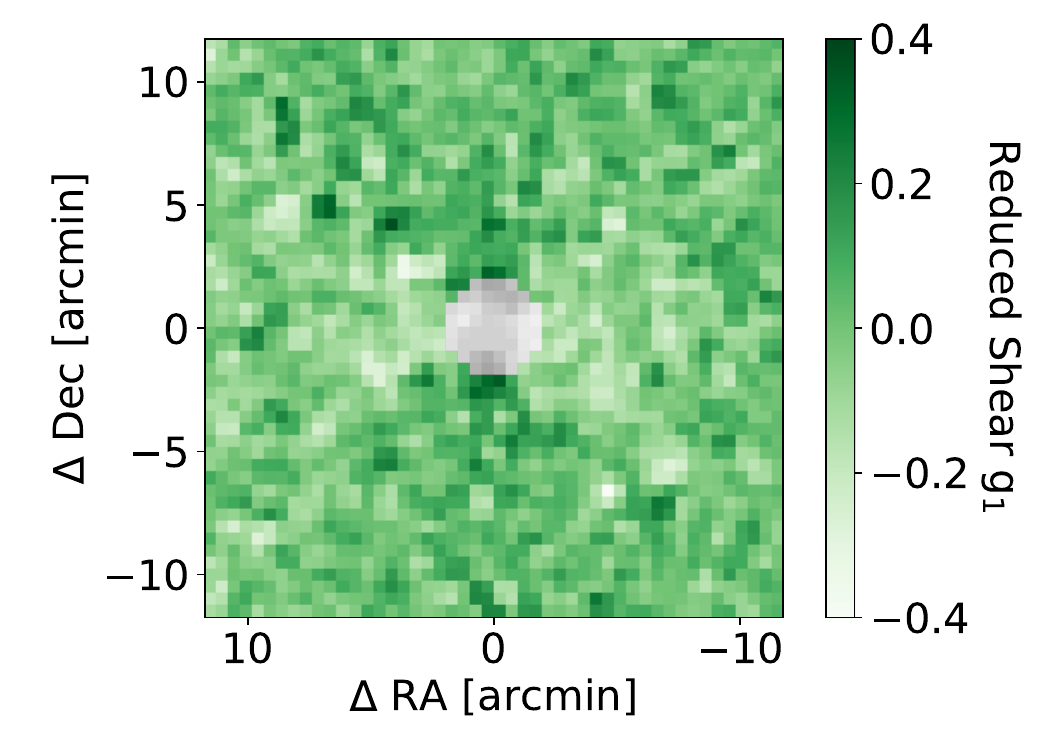}
    \includegraphics[width=0.32\linewidth]{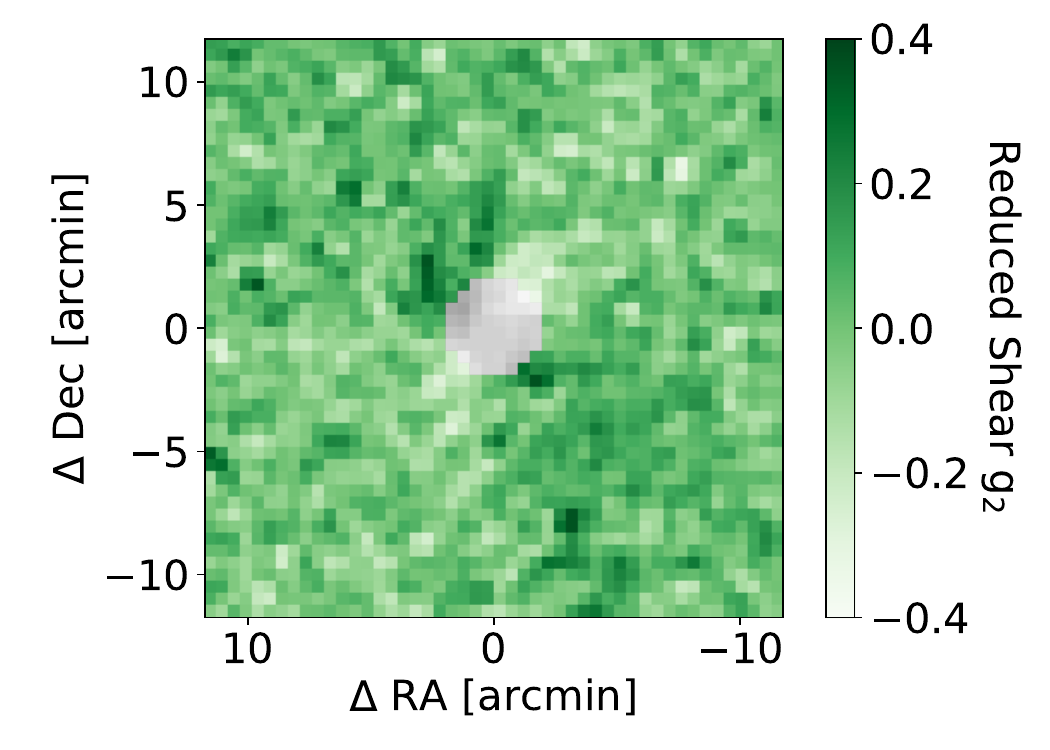}
     \includegraphics[width=0.32\linewidth]{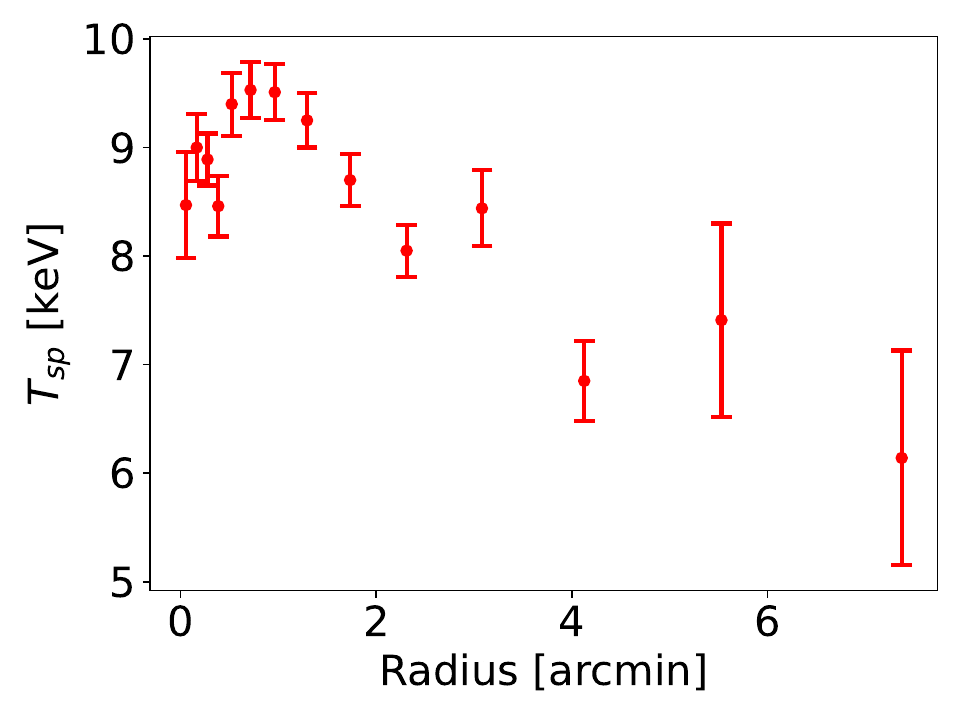}
    \caption{SZ, X-ray and WL data for the CHEX-MATE galaxy cluster \cluster (Abell 1689). The ACT DR6 Compton-\textit{y} map (top left), the X-ray SB map (top middle), X-ray SB radial profile (top right), two component WL shear maps (bottom left and middle), and radial profile of X-ray temperature are shown. The grayed out portions of the maps are data that are excluded from the fit. For the SZ and X-ray portion of the analysis, the data is limited to within the value of \rfive\ obtained by \citet{Planck2016_SZ2}, equal to 7.42$'$ (1.37 Mpc). The X-ray SB employs a one-dimensional and two-dimensional approach. We use the two-dimensional data within the circular region that encloses 80~per cent of the emission and one-dimensional data is used in the radial region outside of this. For the WL shear maps, we exclude data within 300$h^{-1}$ kpc comoving ($\sim$360 kpc physical) to avoid regions containing SL features and significant contamination from cluster member galaxies. The SZ Compton-\textit{y} map is obtained from the combination of observations from ACT and \planck. The two-dimensional X-ray map is a product from \citet{Bartalucci2023} and \citet{Lovisari2024} and was binned using a Voronoi algorithm to ensure constant signal-to-noise. Bright point-like sources in the X-ray map are indicated by white circles and are masked in the analysis. The radial profile of X-ray SB was constructed by \citet{Bartalucci2023}. We chose to use one-dimensional temperature data at all radii. These data were obtained by \citet{Rossetti2024} via spectroscopic fits to the X-ray SB observations. The WL maps were constructed from archival observations from the Subaru Suprime-Cam instrument. }
    \label{fig:ObsData}
\end{figure*}

The two-dimensional X-ray SB map utilized in the two-dimensional portion of the X-ray analysis is produced from \xmm\ observations using the Voronoi tessellation method \citep{Cappellari2003, Diehl2006}. The details of the image production are reported in \citet{Bartalucci2023} and \citet{Lovisari2024}. Bright point sources in the X-ray SB map were masked and excluded from the X-ray analysis. This procedure was performed in a manner to ensure an approximately uniform residual cosmic X-ray background over the entire field of view, sufficiently low to not impact the fit \citep{Ghirardini2019, Bartalucci2023}. The two-dimensional X-ray SB used in this analysis is shown in the top middle of Figure~\ref{fig:ObsData}. Analogous to K24, we also use a one-dimensional projected radial profile of the X-ray SB at larger radii, although we compute this profile in a different manner. In brief, K24 computed the azimuthal mean from the two-dimensional map, while we instead utilize the azimuthal median profile generated from the analysis of \citet{Bartalucci2023}, detailed below.

The radial profile of X-ray SB was constructed by \citet{Bartalucci2023} as follows. First, particle background subtracted Voronoi-binned X-ray SB maps with point sources masked were produced, ensuring 20 counts per bin on average. Concentric annuli centered on the X-ray peak were defined and the SB median profile was extracted by considering in each radial bin the median count rate of the Voronoi cells whose center lies within the annulus. In this way, the use of median profiles limits the bias caused by the emission of sub-clumps and substructures too faint to be identified and masked \citep[e.g.,][]{Roncarelli2013, Zhuravleva2013}. The sky background was estimated also using the median count rate. Then, the sky background subtracted profiles were re-binned to have at least nine counts per bin after background subtraction was performed on the particle background subtracted profiles. The portion of this radial profile used in this analysis is show in the top right of Figure~\ref{fig:ObsData}. 

We note that in this work, we have chosen to use the one-dimensional projected X-ray temperature profile, which differs from the approach of K24, who utilized a two-dimensional temperature map in their analysis. More details regarding this change are outlined in Appendix~\ref{sec:appen:K24_compare}. The technique to generate the projected temperature profiles is detailed in \citet{Rossetti2024}, and we summarize it here. First, spectra in 15 concentric annuli around the X-ray peak were extracted, using the \verb|mos-spectra| tool within the \xmm\ Science Analysis System (SAS) and the Extended Source Analysis Software \citep[ESAS,][]{Snowden2008, Kuntz2008} embedded within SAS, which ensured a roughly constant signal-to-noise ratio in each radial bin. For each region, the corresponding redistribution matrix files (RMF) and ancillary response files (ARF) were extracted using the ESAS tool \verb|pn-spectra|. A spectrum for the cosmic-ray induced particle background (CRPB) was then produced using the \verb|mos-back| and \verb|pn-back| tools. To calibrate the model of the sky background components, the spectrum obtained from an external annulus between 10.6$'$ and the edge of the field-of-view is utilized. The models of all the background components (quiescent particle background, cosmic X-ray background, Galactic halo, local hot bubble, and residual soft protons) are described in detail in \citet{Rossetti2024}. For each region the spectra of the three detectors are jointly fit using XSPEC \citep{Arnaud1996}, where the parameters of the background model are used as priors to derive the temperature, metal abundance and normalization of the ICM. The resulting temperature profile for Abell 1689 is shown in the bottom right panel in Figure~\ref{fig:ObsData}, where the individual points in each radial bin are the mean of the posterior distribution of temperature values and the error bars are the corresponding 68~per cent credible region. 

We make use of a single SZ Compton-$y$ map obtained from the combination of observations from the ground-based Atacama Cosmology Telescope (ACT) and \planck, which was generated by the ACT collaboration and is shown in the top left of Figure~\ref{fig:ObsData}. In particular, we utilize the most recent version of this map, from ACT data release 6 \citep[DR6,][]{Coulton2024, naess2025_ACTDR6}, which is deeper than the ACT-DR4 map used in by K24. The noise rms for the DR4 \textit{y}-map is approximately $9\times10^{-6}$, while for the DR6 \textit{y}-map it is approximately $5\times10^{-6}$ (for more details see Appendix~\ref{sec:appen:K24_compare}). We do not find any point-like sources, corresponding to either radio or dusty star-forming galaxies, within the region of the map utilized for our analysis, and thus no masking or subtraction of such sources was applied. To estimate the noise covariance, we employ the same approach as in K24 based on the technique described in \citet{Pointecouteau2021}. Specifically, we determine which ACT coverage region (BN, D56, or wide) contains the cluster, mask out all locations of detected clusters from \citet{Hilton2021}, and randomly sample 10,000 maps within this region. Then, the covariance of these maps was computed to estimate the noise covariance matrix corresponding to the given cluster.

\subsection{WL data}
\label{sec:WLdata}
We performed a weak-lensing analysis of Abell 1689 using archival observations from the Subaru Suprime-Cam
instrument. Full details of image reduction, photometry, and shape analysis for a sample comprising 40 CHEX-MATE clusters will be presented in forthcoming publications (Gavazzi et al.; Umetsu et al., in preparation). Our study is based on wide-field imaging in the $B$, $V$, $R_\mathrm{C}$, $i^+$, and $z^+$ bands, obtained with Suprime-Cam on the 8.2~m Subaru Telescope.

In brief, the image processing builds on the \astromatic~software suite\footnote{\url{https://www.astromatic.net}}. In particular, the astrometric solution is based on \scamp~\citep{bertin2006,bertin2010} using Abell 1689 exposures but also archival images taken during the same observing run, with a sliding window of several days around the exposures of interest so that at least 3 exposures constitute the same astrometric instrument. The typical internal accuracy is of order 0.015$''$ based on utilizing the Two Micron All Sky Survey as reference \citep[2MASS,][]{2mass}. \swarp~is used for image coaddition \citep{bertin2002} after the exclusion of non-photometric exposures and those collected in poor seeing conditions. We build a model of the PSF and its spatial variations with \psfex~\citep{bertin2011} for all the exposures in all the bands. A PSF model is also built on the coadded frames. The recovered PSF Full Width at Half Maximum (FWHM) is found to be $0\farcs81$, $0\farcs70$, $0\farcs62$, $0\farcs69$, $0\farcs68$, in the $B$, $V$, $R_\mathrm{C}$, $i^+$, and $z^+$ bands, respectively. Galaxy shapes were measured from the $R_\mathrm{C}$-band data, which provide the highest image quality within the dataset, both in terms of depth and seeing. Source ellipticities were measured using the model-fitting capabilities of \sex, assuming the surface brightness distribution of galaxies can be well approximated by a single Sersic profile. The accuracy of this shape measurement technique was extensively assessed in the GREAT3 challenge as part of the Amalgam team \citep{Mandelbaum2015} and further explored in the context of the preparation of the Euclid mission \citep{EuclidIV}. Residual multiplicative of additive biases are expected to be well below the statistical uncertainties for the WL signal expected from this cluster.

Photometric redshifts (photo-$z$) for individual galaxies were estimated by matching the Subaru $BVR_\mathrm{C}i^+z^+$ photometry to the 30-band photometric redshift (photo-$z$) catalog of the 2 square degree COSMOS field \citep{laigle16}. We first degrade the COSMOS2015 photometry when necessary to match the depth of our Abell 1689 data by adding extra noise in the COSMOS2015 catalogue. Then, nearest neighbors in the matched $BVR_\mathrm{C}i^+z^+$ magnitude space are used for each source in order to build a photometric redshift distribution based on 100 nearest neighbors using a fast kd-tree structure \citep{kdtree2}. The resulting mean, median and dispersion redshift are obtained along with a marginalized mean and standard deviation of the distance ratio $\beta \equiv D_{\text{LS}}/D_{\text{S}}$.

The reduced shear signal was measured from a background galaxy sample selected based on a color--color (CC) cut method. This method has been calibrated against evolutionary color tracks of galaxies and photo-$z$ catalogs from deep multiwavelength surveys such as COSMOS \citep[for details, see][]{Medezinski+2010,Medezinski2018src}. In our analysis, CC cuts were performed using $BR_\mathrm{C}z^+$ photometry from Subaru/Suprime-Cam, which offers broad coverage across the optical wavelength range and is well-suited for cluster weak-lensing studies \citep[e.g.,][]{Umetsu2014,Umetsu2022}. These CC cuts yield a background sample of 16,428 galaxies, corresponding to a mean surface number density of $n_g\approx 15$~galaxies~arcmin$^{-2}$. For the selected background sample, we find a weighted mean of $\langle \Sigma_\mathrm{crit}^{-1}\rangle^{-1}=3.53\times 10^{15}$~\msun~Mpc$^{-2}$ and $f_l = \langle\Sigma_\mathrm{crit}^{-2}\rangle/\langle \Sigma_\mathrm{crit}^{-1}\rangle^2 \approx 1.022$.

\begin{figure*}
    \centering
    \includegraphics[width=0.45\linewidth]{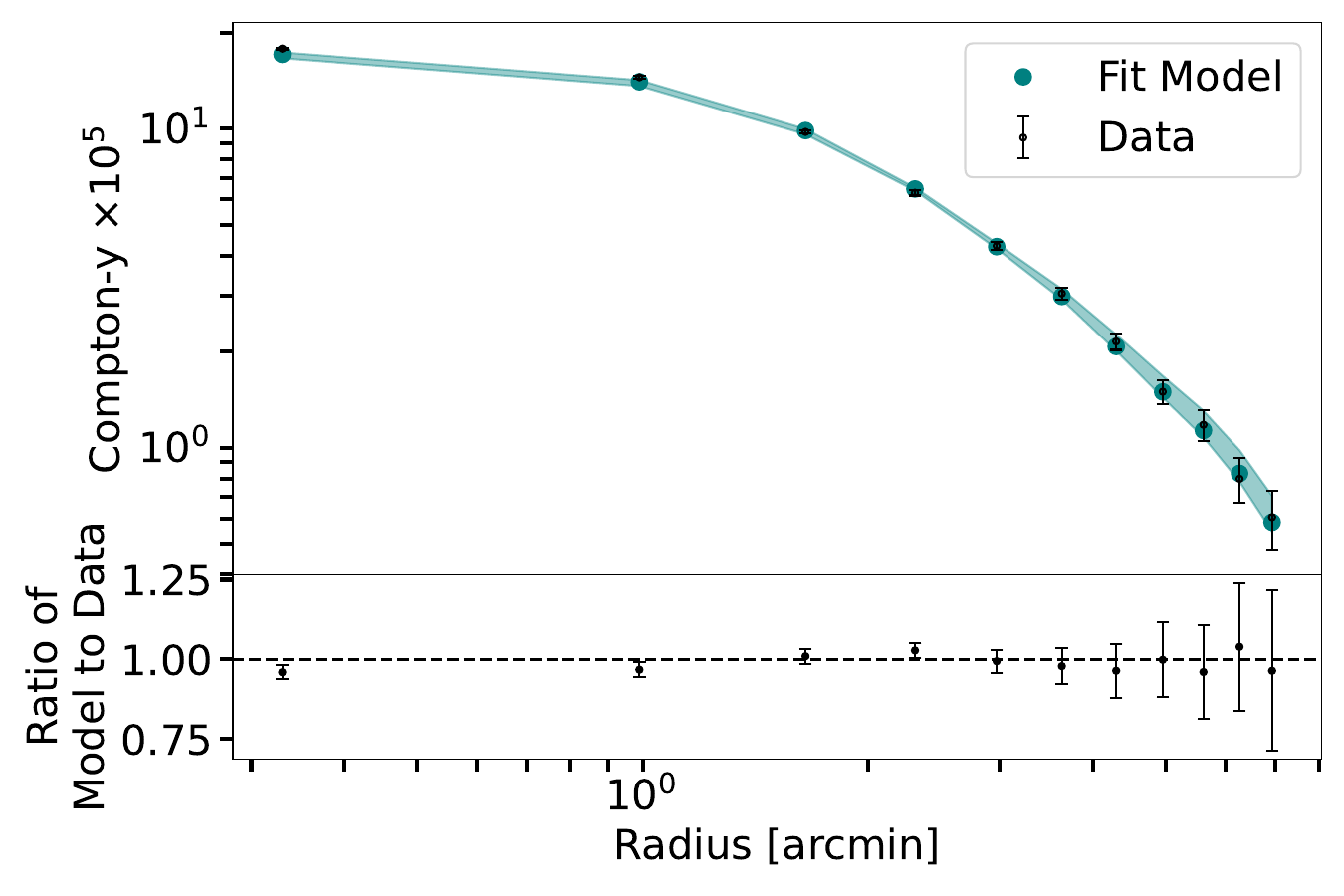}
    \includegraphics[width=0.45\linewidth]{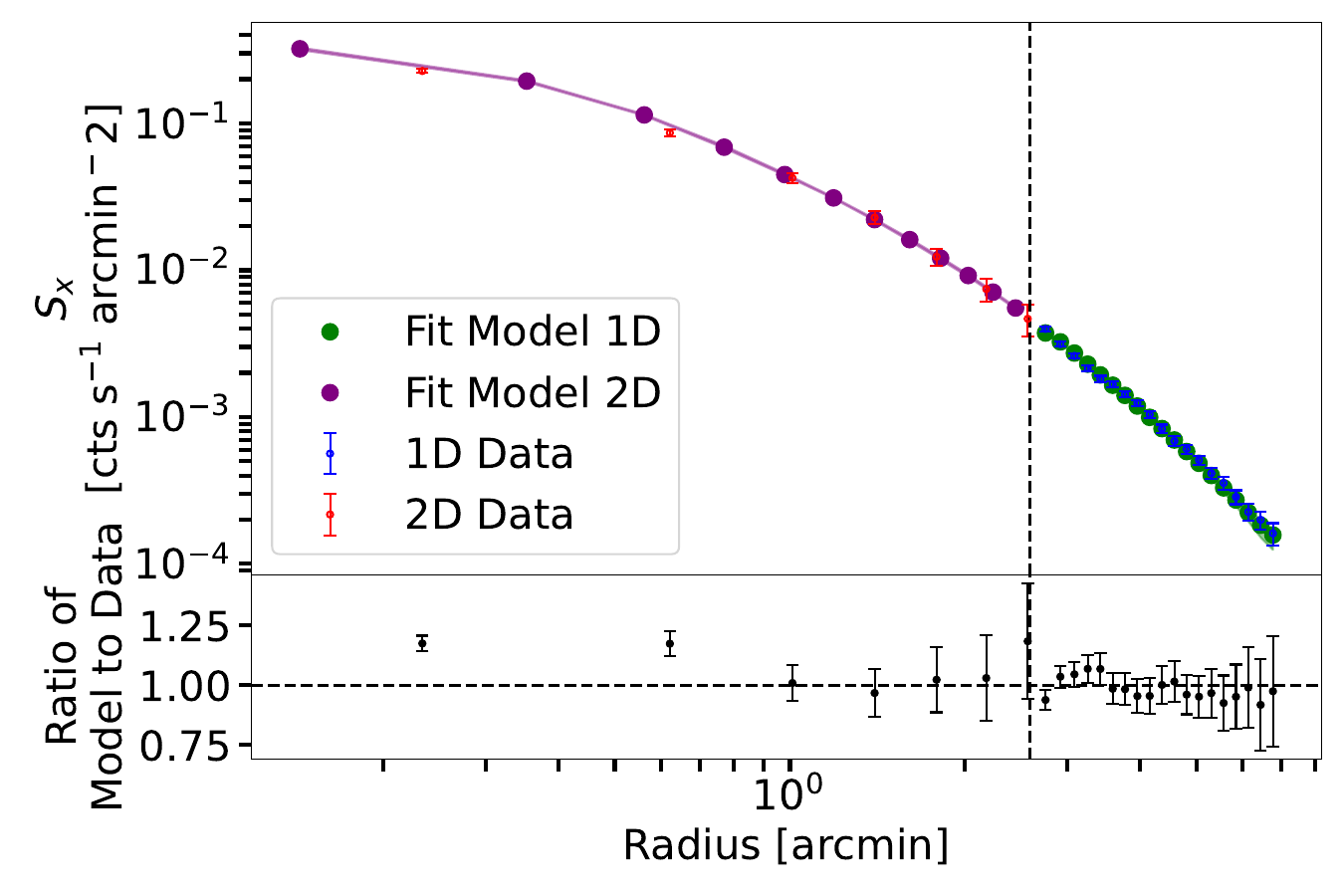}
    \\
    \includegraphics[width=0.45\linewidth]{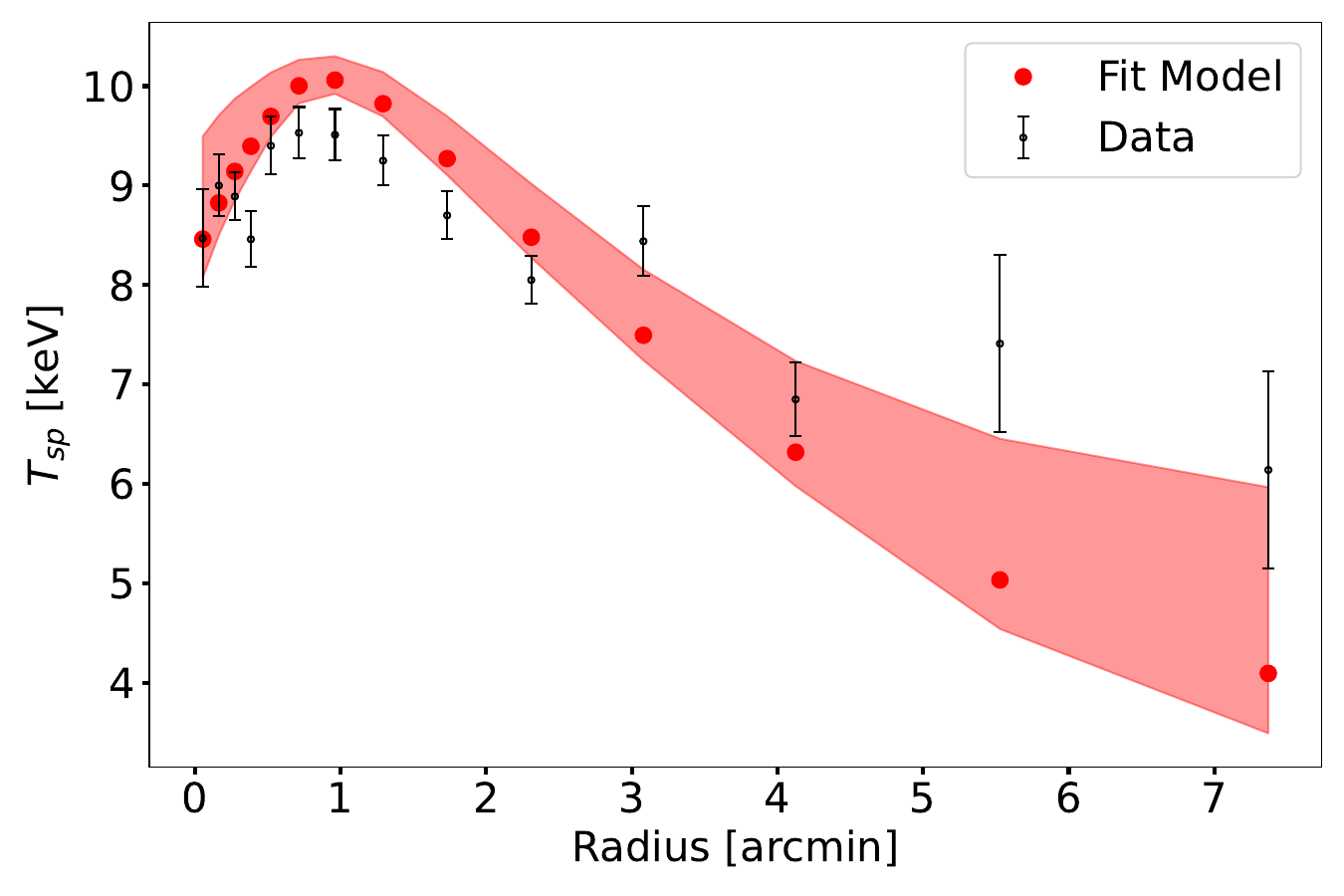}
    \includegraphics[width=0.45\linewidth]{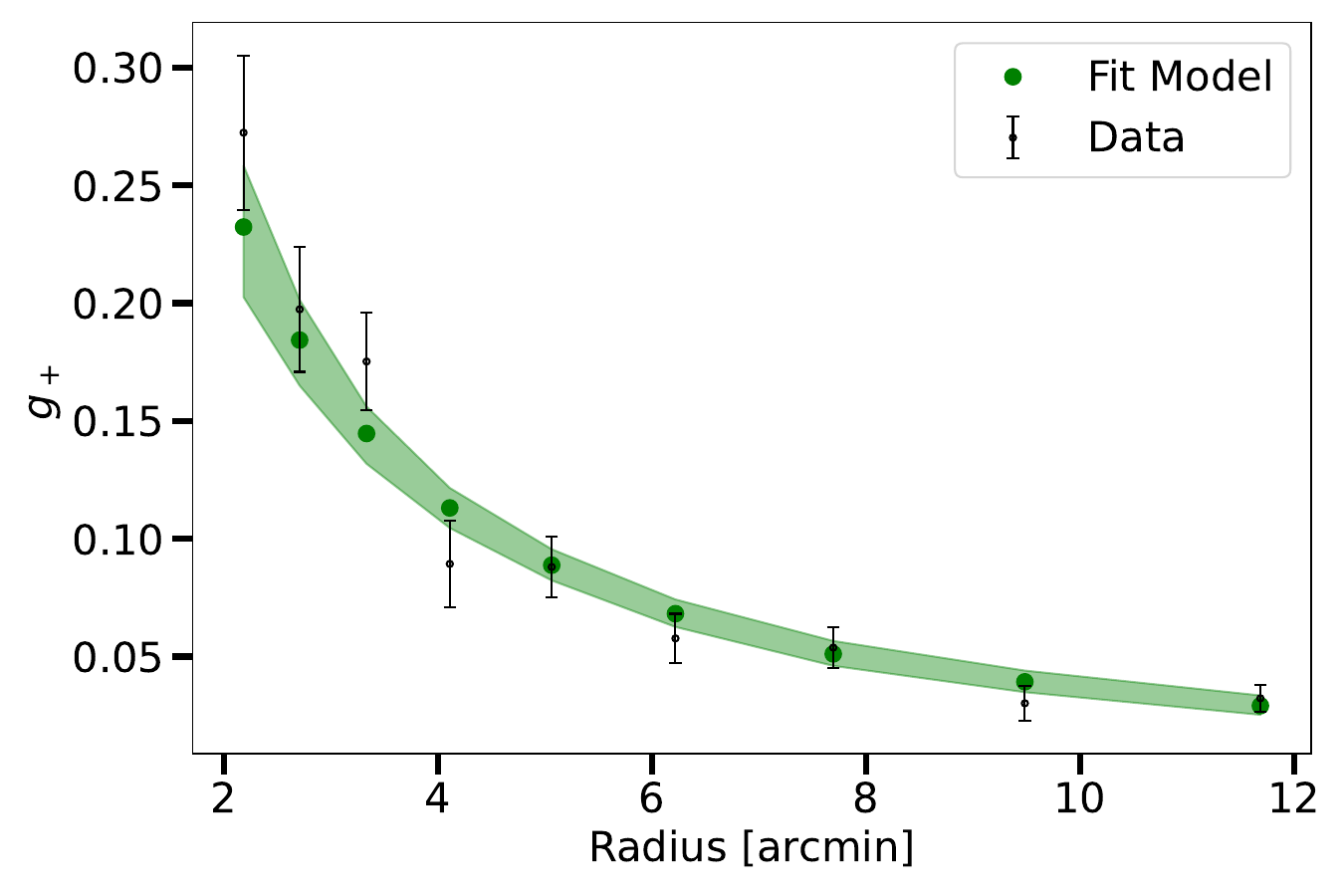}
    \caption{Radial profiles of the model SZ (top left), X-ray SB (top right), X-ray temperature (bottom left) and tangential shear (bottom right) obtained from the accepted samples from the MCMC. For the two dimensional SZ and X-ray model maps, the radial profiles were obtained via azimuthal averaging. A dashed black vertical line on the X-ray SB profile indicates the radius at which the the two-dimensional analysis (left) transitions to a one-dimensional analysis (right). The model tangential shear radial profile was calculated from the two-component model shear maps according to Eq.~\ref{eq:gt_avg}. Radial profiles of the corresponding observational data are plotted alongside the models and were calculated in an analogous manner, aside from the tangential shear data which was calculated directly from the shear catalog and not from our observed two dimensional reduced shear maps. The SZ and X-ray SB model profiles show close agreement to the observational data, so a ratio of the model and the observational data points is shown in the bottom potion of these plots. Note that the profiles of the ICM observables are truncated at the value of \rfive\ obtained by \citet{Planck2016_SZ2}, while the tangential shear extends past this.}
    \label{fig:radial_profiles}
\end{figure*}

Using the CC-selected sample of background galaxies, we compute the source-averaged reduced shear components, $(g_1(\btheta_n), g_2(\btheta_n))$, on a uniform grid of $48\times 48$ pixels with $\Delta\theta=0.5\arcmin$ spacing, covering a $24\arcmin\times 24\arcmin$ field centered on the X-ray cluster peak, as: 
\begin{equation}
\label{eq:bin_shear} 
g(\btheta_n) = g_1(\btheta_n) + ig_2(\btheta_n)
=
\left[
\displaystyle\sum_{s\in n}
w_{g (s)}g_{(s)}
\right]
\left[
\displaystyle\sum_{s\in n} 
w_{g(s)}
\right]^{-1},
\end{equation}
where $g_{(k)}$ is an estimate of $g(\btheta)$ for the $s$th galaxy at $\btheta_{(s)}$, and $w_{g(s)}$ is its statistical weight, $w_{g(s)} = 1/(\sigma^2_{g(s)}+\sigma^2_\mathrm{rms})$. Here $\sigma_{g(s)}$ represents the measurement uncertainty for $g_{(s)}$, and $\sigma_\mathrm{rms}$ is the rms ellipticity for the background sample, which we take as $\sigma_\mathrm{rms}=0.4$ or equivalently, $0.4/\sqrt{2}\simeq 0.28$ per component \citep[e.g.,][]{Umetsu2014}. The resulting maps for $g_1$ and $g_2$ are shown in the bottom right and middle of Figure~\ref{fig:ObsData}.

The source-averaged expectation (denoted by a hat symbol) for the observed $g(\btheta_n)$ is given by \citep{Seitz+Schneider1997,Umetsu2020rev}
\begin{equation}
\label{eq:g_ave}
\widehat{g}(\btheta_n) \simeq \frac{\gamma(\btheta_n)}{1-f_l \kappa(\btheta_n)}.
\end{equation}
The error variance  $\sigma_{g,n}^2$ for $g_n$ is expressed as
\begin{equation}
 \begin{aligned}
\label{eq:bin_shearvar}
\sigma^2_{g,n}=
 \frac{ \sum_{s\in n} w_{g(s)}^2 \sigma^2_{g(s)} }
{ \left[
 {\sum_{s\in n} w_{g(s)}}
  \right]^{2}}.
  \end{aligned}
\end{equation}

We excise the center of the shear maps up to a radius of 300$h^{-1}$ kpc comoving ($\sim$360 kpc physical) to avoid regions that contain SL features and significant contamination from cluster member galaxies. Because shear is a non-local measurement, we assume that the data are sensitive to the underlying mass distribution outside of approximately half this radius, or 180~kpc. Specifically, with minimal sensitivity to profile shape, approximately half of the shear signal measured at a given radius is sourced from the mass distribution external to half of that radius. Applying the same threshold to the extent of the Suprime-Cam field of view, equal to 2.1~Mpc, we assume that the data do not fully constrain the mass distribution beyond approximately 1.05~Mpc. 
\section{Results} \label{sec:Results}
The triaxial analysis technique was applied using the datasets described above for Abell 1689. For this fit, we align the model center with the X-ray peak \citep{Bartalucci2023}, which is consistent with the position of the BCG \citep{Sereno2013}. We note that for a morphologically regular cluster, like Abell 1689, such co-alignment of the different probes is expected, and thus our exact choice of center is not expected to impact the overall model fit. Furthermore, since this cluster is elongated toward the observer, any offset will be preferentially oriented in that direction, thus resulting in a relatively small POS offset. Figure~\ref{fig:A1689_cornerplot} shows the posterior distributions of the model parameters that describe the triaxial fit of Abell 1689. These distributions were constructed from 45 Markov chains of length 20,000 steps using a lower limit on the acceptance rate of 15~per cent. The first 10,000 steps of each chain were discarded. We find that the priors are informative for two parameters: $\cos\theta$ is limited by the physically-allowed limit that its value is $\leq$1, and \ctwoc\ is influenced by our prior based on the simulations of \citet{Diemer2019}. Overall, we can conclude that our parameter constraints are not prior-limited. 

We find axial ratios of $q_{\text{ICM,1}} = 0.75 \pm 0.01$ and $q_{\text{ICM,2}} = 0.82 \pm 0.01$. The first axial ratio is larger than the one found in K24, equal to $0.65 \pm 0.02$, while the second axial ratio is consistent with their value. Our fit indicates that the major axis of Abell 1689 is almost perfectly aligned with the line of sight, with cos$\theta = 0.99 \pm 0.01$, in agreement with the K24 result. We find an elongation of $\mathcal{R}_{LP} = 1.27 \pm 0.02$, which, when converted to the definition for elongation in K24, gives $e_\parallel = 1.21 \pm 0.02$, consistent with the value found in that work. We refrain from directly comparing the individual parameters of the electron density and pressure profiles, as strong degeneracies between many of these parameters make such comparison difficult. However, we do compare radial profiles of the electron density and pressure from the two fits, and, as shown in Figure~\ref{fig:eprofiles_compare}, they are in good agreement. In aggregate, other than the first axial ratio, this implies that the geometric and gas properties of Abell~1689 are not strongly impacted by our changes to fitting algorithms and observational datasets. In Appendix~\ref{sec:appen:K24_compare} we list the specific updates to the fitting algorithms and observational datasets relative to K24, along with the relative impact of each of these updates on the above fitted values. 

Figure~\ref{fig:radial_profiles} shows the fitted radial profiles reconstructed from the parameter posteriors. For the two-dimensional data, these were calculated via azimuthal averaging of the two-dimensional model and observed maps. We see that for the Compton-\textit{y} and X-ray SB profiles, the model follows the the data closely, although there is some indication of the model over-estimating the X-ray SB signal at the smallest radii. The recovered X-ray temperature profile differs from the observed data at the $\sim 1\sigma$ level at most radii and is generally slightly above the observed data at small radii and slightly below the observed data at large radii. This is likely due to the available degrees of freedom in our model. Two independent thermodynamic profile shapes (density and pressure) must simultaneously describe three observed quantities (X-ray SB and temperature, along with SZ brightness), with $\mathcal{R}_{LP}$ solely impacting the relative normalization. As a result, the higher signal-to-noise SZ data primarily constrain the shape of the pressure profile, which is slightly discrepant with the observed X-ray temperature profile shape. However, the fitted value of $\mathcal{R}_{LP}$ ensures that the overall normalization of the fitted temperature profile is in good agreement with the observed data. Additionally, we show the tangential shear ($g_+$) obtained from our fit and calculated using Eq.~\ref{eq:gt_avg} compared to the observed values of $g_+$ derived directly from a two-dimensional azimuthal averaging of the shear catalog. These values were obtained from \citet{Loris2025} and were not calculated directly from the observed two-dimensional component reduced shear maps shown in Figure~\ref{fig:ObsData}. We see good agreement between our best fit model $g_+$ and the observed data at all radii.

\begin{figure}
    \centering
    \includegraphics[width=0.95\linewidth]{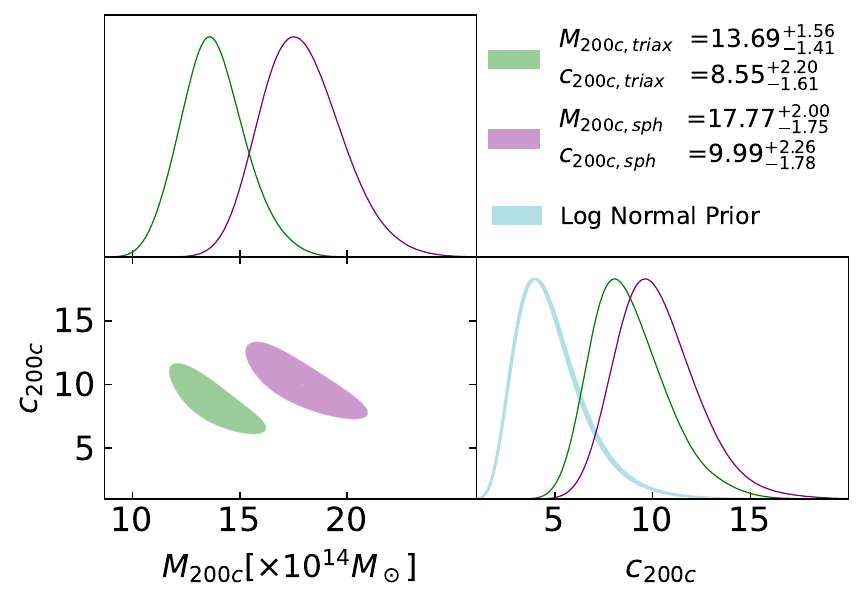}
    \caption{Posterior distributions of the mass and concentration obtained for a fit to Abell 1689 using SZ, X-ray SB and temperature, and WL data assuming a triaxial geometry (in green, labeled as M$_{200c,\text{triax}}$ and $c_{200c,\text{triax}}$) and using only WL data, assuming a spherical geometry (in purple, labeled as M$_{200c,\text{sph}}$ and $c_{200c,\text{sph}}$). The 68~per cent credible regions for the two results are color-filled in the bottom left subplot. The log normal prior we apply to \ctwoc, which depends on mass, is marginalized over the 68~per cent credible bounds of M$_{200c,\text{triax}}$ and plotted in blue in the lower right subplot. Note that the prior is strongly informative and the upper edges of the posterior distributions for \ctwoc\ are largely set by it.}
    \label{fig:WL_sph_triax}
\end{figure}

For the total matter density profile, we recover \mtwoc\ $= 13.69_{-1.41}^{+1.56}$ $[10^{14}$\msun] and \ctwoc$= 8.55_{-1.61}^{+2.20}$. To gauge any potential bias introduced by projection effects, we also performed a spherical fit ($q_{\text{ICM,1}} = q_{\text{ICM,2}} = 1$) to the WL data within our pipeline to obtain $\text{M}_{200c,\text{sph}} = 17.77_{-1.75}^{+2.00}$ [$10^{14}$\msun] and $c_{200c,\text{sph}} = 9.99_{-1.78}^{+2.26}$. We show the posterior distributions of these triaxial and spherical values in Figure~\ref{fig:WL_sph_triax}. Prior fits of spherical models have suggested a surprisingly high concentration for this system, consistent with our value of $c_{200c,\text{sph}}$, and triaxiality and projection have been proposed as a possible explanation for such values \citep{Broadhurst2008, Oguri2009, Sereno2010, Meneghetti2014}. However, the large value of \ctwoc\ is retained in our triaxial fit, suggesting that it may be intrinsic to the system rather than a result of projection. We also note that the informative log-normal prior may actually bias our result to a lower value of \ctwoc, see Section~\ref{sec:MockData}, thus suggesting that the true concentration may be even higher. We see that the right edge of the posterior distribution for \ctwoc\ in Figure~\ref{fig:WL_sph_triax} is set by the shape of the right edge of the prior. Performing the fit with  flat prior ($\mathcal{U}(0, 20)$) on \ctwoc\ returns a posterior distribution for this value that hits the upper end of the prior. Additionally, the exceptionally high concentration obtained in our fits aligns with known selection effects. That is, strong-lensing clusters tend to be biased toward compact, LOS-aligned configurations, like Abell 1689. The much higher mass obtained in the spherical fit is in agreement with our expectations from the fitted geometry, given the LOS elongation $\mathcal{R}_{LP} = 1.27$.
 
\begin{figure}
    \centering
    \includegraphics[width=0.95\linewidth]{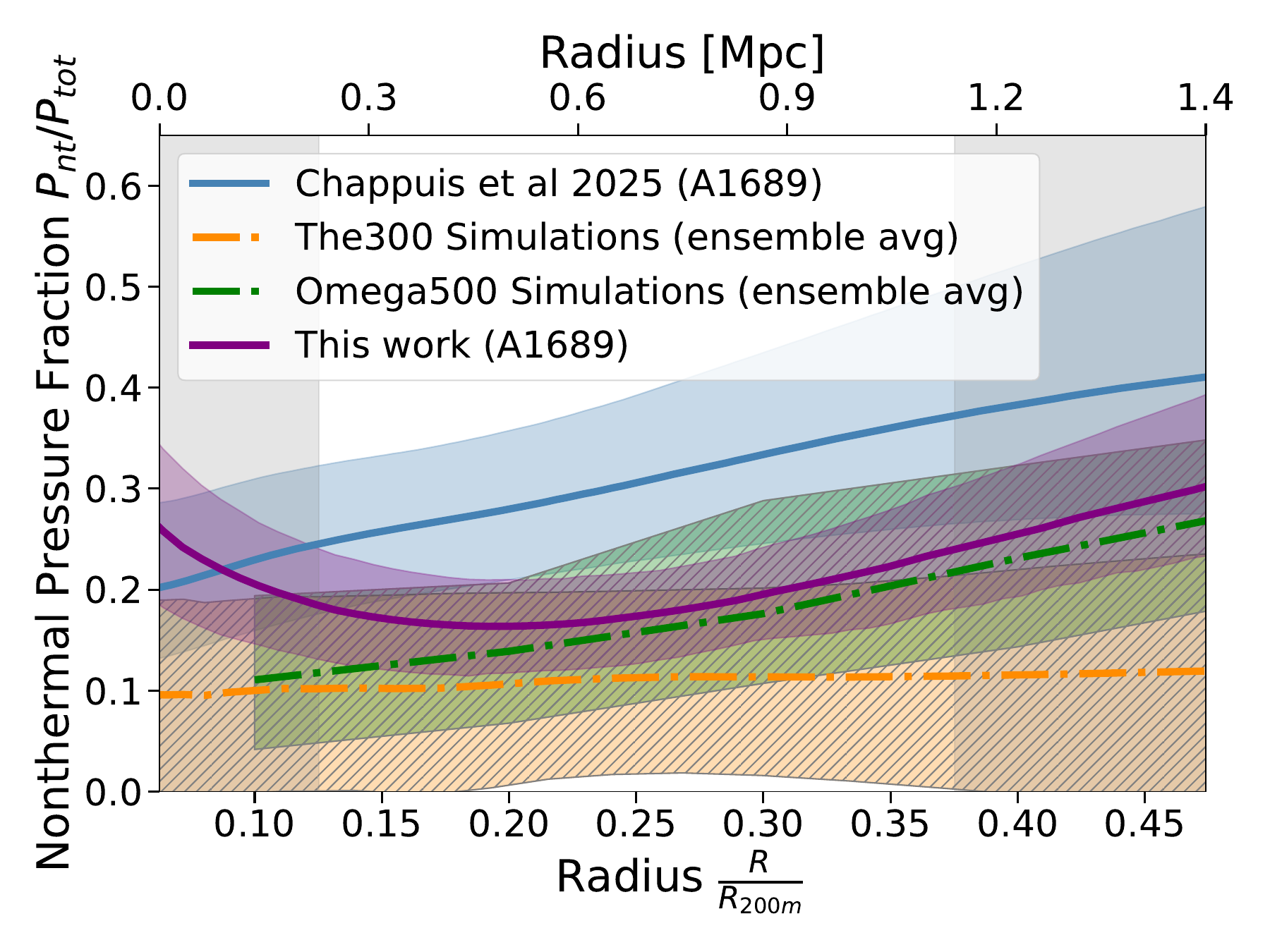}
    \caption{The non-thermal pressure fraction (\pnt/\ptot) from our fit to Abell 1689, along with the corresponding 68~per cent credible region (shaded purple), calculated from the accepted MCMC samples as outlined in Section~\ref{sec:Pnt}. For comparison, the \pnt/\ptot\ result obtained for the same cluster by \citet{Loris2025} is shown in blue, also plotted with the corresponding 68~per cent credible region. The results in yellow and green show the ensemble-average \pnt/\ptot\ result from independent analyses of simulated clusters from The300 \citep{Sayers2021} and Omega500 \citep{Nelson2014} simulations. The shaded hashed regions for these correspond to the 68~per cent intrinsic cluster-to-cluster scatter within those simulations. Shaded gray denotes regions where the mass distribution may not be fully constrained from the observed WL shear signal (see Section~\ref{sec:WLdata})}
    \label{fig:Pnt} 
\end{figure}

As discussed in Section~\ref{sec:Pnt}, the calculation of the non-thermal pressure fraction was performed using the chains of accepted samples from the MCMC for the individual parameters. Figure~\ref{fig:Pnt} shows a radial plot of the recovered non-thermal pressure fraction \pnt/\ptot\ with the 68~per cent credible region in purple. The radial range of this profile was set on the lower end by the WL data ($\sim$180 kpc, see Section~\ref{sec:WLdata}) and on the upper end by the maximum radius considered for the gas observables, (1.37 Mpc, see Section \ref{sec:ne&Pe}). We find that \pnt/\ptot\ has a value of $\sim$25~per cent at the minimum considered radius, decreases to a minimum value of $\sim$18~per cent near 600~kpc, and then gradually increases to $\sim$30~per cent at the maximal considered radius. Included in the plot is the non-thermal pressure fraction and 68~per cent credible region obtained by \citet{Loris2025}, found from their analysis of this same cluster. Additionally, we plot the non-thermal pressure fraction profiles from two independent analyses of an ensemble of simulated galaxy clusters in yellow \citep[][The300]{Sayers2021, Cui2018} and green \citep[][Omega500]{Nelson2014}, along with their respective intrinsic cluster-to-cluster scatter. 

\begin{figure}
    \centering
    \includegraphics[width=0.9\linewidth]{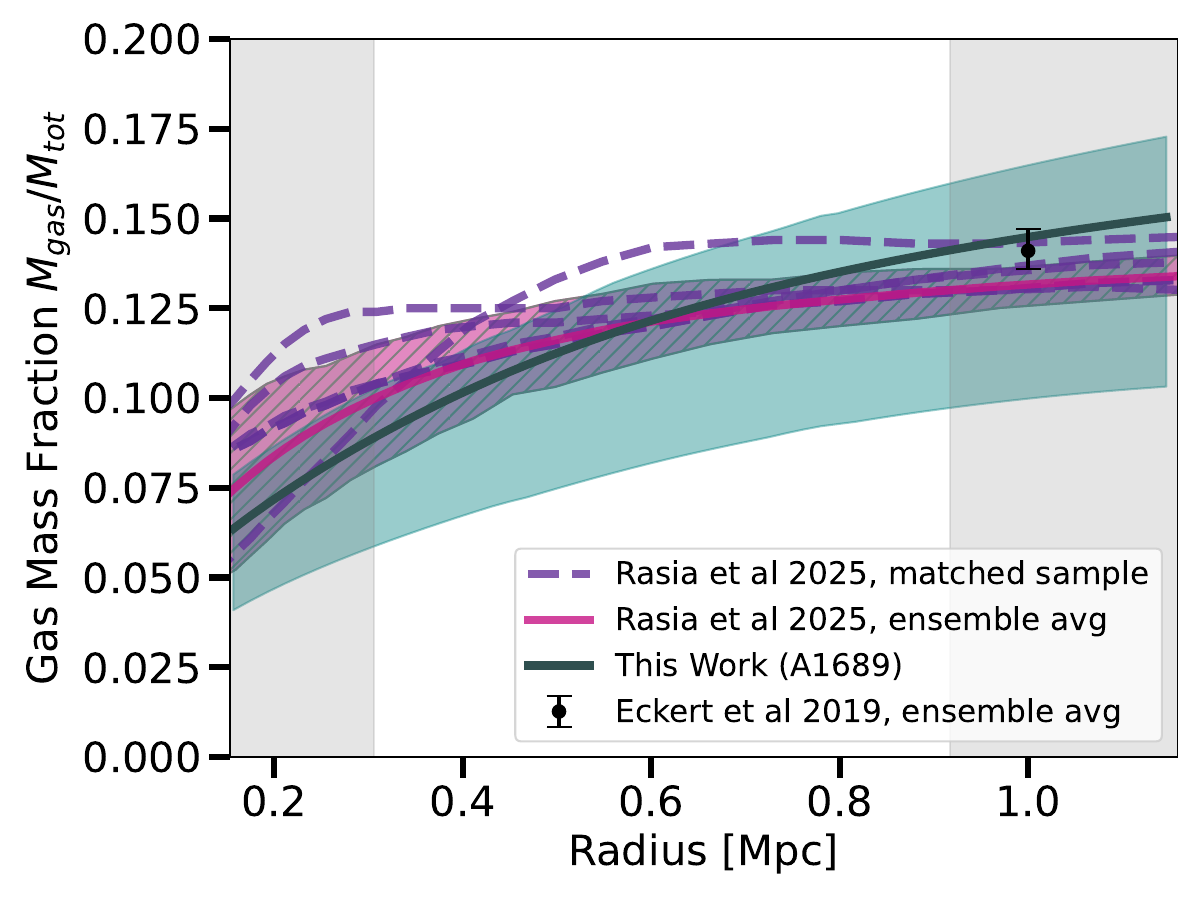}
    \caption{Gas mass fraction computed from our fits to the gas and total mass density compared to the ensemble average profile obtained from simulated galaxy clusters from The300 used in the analysis of \cite{Rasia2025}. The 68~per cent credible region and intrinsic scatter are shown in the shaded region of the respective profiles. We also show a matched sample of five simulated galaxies clusters used in the analysis of \citet{Rasia2025} that have a similar mass and concentration relation to Abell 1689. Additionally, we compare to the result from \citet{Eckert2019}, who provide the median value of the gas mass fraction obtained from a sample of 12 observed galaxy clusters at $R_{500c}$. Shaded gray denotes regions where the mass distribution may not be fully constrained from the observed WL shear signal (see Section~\ref{sec:WLdata})}
    \label{fig:mass_frac}
\end{figure}

We show in Figure~\ref{fig:mass_frac} the gas mass fraction calculated from our fit to the electron density, converted to a gas density using Eq.~\ref{eq:gas_density}, and the total density which we model as an NFW profile. From these, we obtain the three-dimensional integrated gas mass and total mass as functions of radius, and their ratio provides the gas mass fraction. We compare our profile to that obtained from the ensemble average profile of simulated clusters from the Three Hundred Project \citep[][The300]{Cui2018} used in the analysis of \citet{Rasia2025} and we find good agreement at all radii. The shaded hashed region of the pink curve is the 68~per cent intrinsic cluster-to-cluster scatter on the ensemble average. We also show the gas mass fraction of five simulated clusters from the sample used in \citet{Rasia2025} that have similar mass-concentration relations (dashed purple lines) as what we find for Abell 1689. We note that our measured gas mass fraction is also consistent with that found in a recent observational study from \citet{Eckert2019}.

\section{Discussion} \label{sec:Discussion}
K24 provides a detailed comparison of the axial ratios and inclination angle obtained from their triaxial fit to those found in other studies of Abell 1689 \citep{Sereno2012, Umetsu2015}. Because our results on the triaxial geometry are broadly in agreement with K24, we instead focus our discussion on comparing our total mass and non-thermal pressure results with those in the literature. In particular, we note that results from \citet{Loris2025} are particularly relevant. They use identical one-dimensional observational datasets for the the X-ray temperature and X-ray SB as used in this work, though their X-ray SB analysis employs this one-dimensional data at all radii. Additionally, we have confirmed that the azimuthally-averaged SZ and X-ray SB data shown in Figure~\ref{fig:radial_profiles} are consistent with the corresponding one-dimensional projected profiles for those data used in \citet{Loris2025}, which were derived using identical datasets from \xmm\ and ACT DR6. We further note that the same underlying shape catalogs from Subaru were utilized in our WL analyses. Although, as noted above, \citet{Loris2025} considered the one-dimensional tangential shear profile while we fit the two-dimensional two-component shear maps. With regard to overall three-dimensional geometry, \cite{Loris2025} find a LOS elongation of $1.30\pm0.03$, consistent with our value of $\mathcal{R}_{LP} = 1.27 \pm 0.02$. 

\renewcommand{\arraystretch}{1.5}
\begin{table*}[!t]
\caption{\label{tab:cM_results} Published Mass and Concentration Measurements of A1689}
    \centering
    \begin{tabular}{c|c|c|c|c}
    \hline\hline 
    Author & \mtwoc\ [$10^{14}$\msun] & \ctwoc & Prior on \ctwoc & Data \\ 
    \hline
    \citet{Oguri2005} & $16.3_{-5.1}^{+2.6}$ & $13.6_{-1.5}^{+1.8}$  & Flat & WL/SL \\
    \citet{Sereno2011} & $13 \pm 2$ & $\phantom{1}7.3 \pm 0.8$ & Flat & WL/SL \\
    \hline
    \citet{Morandi2011} & $25.9 \pm 0.9$ & $\phantom{1}5.7 \pm 0.5$ & Flat & X-ray/WL/SL \\
    \citet{Sereno2013} & $13.3 \pm 1.7$ & $\phantom{1}7.8 \pm 0.7$ & Flat & X-ray/SZ/WL/SL \\ 
    \citet{Umetsu2015} & $17.8 \pm 2.3$ & $\phantom{1}8.4 \pm  1.3$ & Flat & X-ray/SZ/WL/SL \\
    \hline 
    \citet{Loris2025} & $17.6_{-1.5}^{+1.7}$ & $\phantom{1}6.4_{-0.6}^{+0.8}$ & Flat & X-ray/SZ/WL \\ 
    This work & $13.7_{-1.4}^{+1.6}$ & $8.6_{-1.6}^{+2.2}$ & Log Normal & X-ray/SZ/WL \\ 
    \hline
    \hline
    \end{tabular}
    \label{tab:compare_cM}
    \vspace{2pt}
\end{table*}

\subsection{Mass and concentration} 
Similar analyses have been performed on Abell 1689 to obtain constraints on \mtwoc\ and \ctwoc\ in both a spherical and triaxial basis. Our spherical mass and concentration results are consistent with the previous WL analysis of Abell 1689 by \citet{Umetsu2008}, which found \mtwoc\ $=17.6_{-2.0}^{+2.0} \times 10^{15}$\msun\ and \ctwoc$=10.7_{-2.7}^{+4.5}$, using WL shear and magnification data. Our results are also in agreement with the mass and concentration values found from a spherical WL-only analysis by \citet{Loris2025}. 

A summary of the results from previous analyses employing triaxial mass models are shown in Table~\ref{tab:cM_results} and described in detail below. In aggregate, we generally find excellent agreement with these prior studies.
\begin{itemize}
    \item \citet{Oguri2005} fit a triaxial NFW profile to a combined SL and WL dataset from \hst\ and Subaru. They produce a model lensing convergence field for a triaxial halo that is specified by 7 parameters, which are the virial radius of the triaxial halo, the concentration, two axial ratios and three Euler angles. Flat priors are applied to all parameters. The virial mass is not a free parameter in the fit, but is derived from the model parameters. From their results, we calculate NFW-equivalent triaxial model parameters of \mtwoc\ $=16.3_{-5.1}^{+2.6}\times10^{14}$~\msun\ and \ctwoc$=13.6_{-1.5}^{+1.8}$ \citep{Umetsu2015}. Their \mtwoc\ value is fully consistent with ours and our \ctwoc\ values differ at a significance of ~$2\sigma$. 
    \item \citet{Sereno2011} fit a triaxial NFW profile to a combined set of SL and WL data. The WL maps included shear plus magnification based on Subaru images and SL multiple image systems were derived from \hst. Their model parameters and priors were identical to those fit by \citet{Oguri2005}, and they found \mtwoc\ $=(13\pm2)\times10^{14}$~\msun\ and \ctwoc$=7.3\pm0.8$. These values are fully consistent with our results. They also performed a WL-only triaxial fit which utilized priors on geometry from N-body simulations to find $(13\pm4)\times10^{14}$~\msun\ and \ctwoc$=10\pm3$, again, fully consistent with our results.
    \item \citet{Morandi2011} performed a joint fit to X-ray (\textit{Chandra}), WL (Subaru), and SL (\hst) measurements. They assume a gNFW triaxial model of the DM, with free parameters being the concentration, scale radius, the inner slope of the total matter profile, and two axial ratios. From their fit to this profile they find \mtwoc\ $= (25.9\pm0.9)\times10^{14}$~\msun\ and \ctwoc$= 5.71\pm0.47$. There is significant discrepancy between these values, our results, and other results found in the literature. However, we note that the deviation is closely aligned with the strong degeneracy direction between concentration and mass, suggesting that the true discrepancy is much lower than what is implied by the marginalized one-dimensional parameter values.
    \item \citet{Sereno2013} utilized a multi-probe data set comprised by a two-dimensional WL convergence map from Subarau, along with SL observations \hst, one-dimensional X-ray SB measured by \textit{Chandra}, and one-dimensional X-ray temperature data obtained from spectroscopic fits to \textit{XMM-Newton} data. Additionally, they consider the total integrated SZ signal obtained from combining data from 7 different facilities. They assume an ellipsoidal NFW profile to describe the density distribution of total matter and fit for the mass, concentration, the two axial ratios of the matter distribution, the minor to major axial ratio of the ICM, and elongation and find \mtwoc\ $= (13.3\pm1.7)\times10^{14}$~\msun\ and \ctwoc$= 7.8\pm0.7$. These values are again consistent with our results.
    \item \citet{Umetsu2015} combined WL data from Subaru with SL from \hst, X-ray observations from {\it Chandra} and interferometric ground-based SZ data. Note that the WL data used in this work is deeper than that used in \citet{Sereno2013}. They construct the projected matter distribution from a joint WL analysis of two-dimensional shear and azimuthally integrated magnification constraints, assuming an elliptical NFW profile. Uniform priors are applied to the free parameters, which are the same as the above analysis. For their full multi-probe analysis, they find \mtwoc\ $= (17.8 \pm 2.3) \times 10^{14}$~\msun\ and \ctwoc$=8.36\pm1.27$, consistent with our findings.
\end{itemize}

While they did not perform a full three-dimensional triaxial analysis, we also compare with the findings of \citet{Loris2025}, who, as noted earlier, utilize SZ, X-ray, and WL data products derived from the same pipeline as our work \citep{chexmate2021}. 
They find \mtwoc\ $= 17.6_{-1.5}^{+1.7} \times 10^{14}$~\msun\ and \ctwoc$= 6.44_{-0.56}^{+0.77}$, which are $\sim 1.6\sigma$ higher and within $1\sigma$, respectively, relative to our results. Their model assumes a generalized HSE relation within the fit, allowing the ICM gas model to be connected to the total matter density model. This means at small radii the X-ray and SZ data provide a constraint on the total mass profile, and thus, \ctwoc. Compared to our results, along with those based on combining WL and SL, their result suggests that the gas probes favor a shallower inner mass profile than gravitational lensing. Prior results also imply that including SL data yields lower values of \ctwoc. Furthermore, as with \citet{Morandi2011}, our mass and concentration values differ along the primary degeneracy direction, suggesting that the true discrepancy in our results is less than what is implied by the one-dimensional marginalized parameter values. Finally, we note that their pipeline is able to obtain more precise constraints on the mass profile, primarily for \ctwoc, as it was developed and optimized for such measurements with the inclusion of constraints from the gas data. 

\subsection{Non-thermal pressure fraction}
The level of non-thermal pressure support in Abell 1689 has been addressed in two previous studies, both of which employed the same generalized HSE equation as in our work to describe the thermal state of the cluster (i.e., Eq. \ref{eq:gHSE}). These studies also assume a triaxial basis. Specifically, \citet{Sereno2013} found the contribution to non-thermal pressure in Abell 1689 is $\sim$20--50~per cent in the inner regions ($\lesssim$ 300 kpc) and $\sim 25 \pm 5$~per cent at $\sim$1.5 Mpc. \citet{Morandi2011} assumed a model of the non-thermal pressure of the gas that is a constant fraction of the total pressure and found a contribution on the order of $\sim 20$~per cent. Our derived non-thermal pressure support is consistent with these results, with a higher level of precision corresponding to a 68~per cent credible region with a total spread of approximately 10~per cent across the full radial range.   

\citet{Loris2025} found a non-thermal pressure fraction equal to $\sim$20~per cent at small radii and increasing steadily to $\sim$ 40~per cent near the maximum radius considered in our analysis (see Figure~\ref{fig:Pnt}). This result is consistent with ours at the 1$\sigma$ level over the full radial range. The modest differences in radial shape between the profiles are a result of modeling differences. \citet{Loris2025} uses a parametric form of \pnt\ within their fit, constraining the overall shape of the profile. In our analysis the \pnt/\ptot\ profile is derived from a combination of the gas density, gas pressure, and total density profiles, and it can thus obtain a wide range of possible radial shapes. We find tighter constraints on our radial profile of the non-thermal pressure fraction than \citet{Loris2025}, which we again attribute to modeling differences.

In Figure~\ref{fig:Pnt}, we also display the non-thermal pressure fraction obtained from large samples of simulated clusters, in particular, The300 and the Omega500 simulations \citep{Nelson2014}. The analysis described in \citet{Sayers2021} was applied to 315 simulated clusters from The300 to obtain the ensemble average profile shown in yellow. The intrinsic cluster-to-cluster scatter for this sample is shown as the hashed yellow region. Over most of the radial range, the 68~per cent credible region from our fit lies above the upper bound of the yellow hashed region, indicating that Abell~1689 has a somewhat higher than typical non-thermal pressure fraction relative to The300 clusters. The non-thermal pressure fraction plotted in green is from an analysis applied to 65 simulated clusters is presented in \citet{Nelson2014} from the Omega500 simulations. The intrinsic cluster-to-cluster scatter for this sample is shown as the hashed green region. For the full radial range considered, the value of \pnt/\ptot\ that we obtain for Abell 1689 is consistent with the ensemble average from the Omega500 clusters.

\begin{acknowledgements}
This research was supported by the International Space Science Institute (ISSI) in Bern, through ISSI International Team project \#565 ({\it Multi-Wavelength Studies of the Culmination of Structure Formation in the Universe}). AG, JS, HS and JK were supported by NASA Astrophysics Data Analysis Program (ADAP) Grant 80NSSC21K1571.  S.E., M.S., M.R., HB, FDL, PM, … acknowledge the financial contribution from the contracts Prin-MUR 2022 supported by Next Generation EU (M4.C2.1.1, n.20227RNLY3 {\it The concordance cosmological model: stress-tests with galaxy clusters}), and from the European Union’s Horizon 2020 Programme under the AHEAD2020 project (grant agreement n. 871158). LL acknowledges the financial contribution from the INAF grant 1.05.12.04.01. M.G. acknowledges support from the ERC Consolidator Grant \textit{BlackHoleWeather} (101086804). MS acknowledges the financial contributions from contract INAF mainstream project 1.05.01.86.10 and INAF Theory Grant 2023: Gravitational lensing detection of matter distribution at galaxy cluster boundaries and beyond (1.05.23.06.17). HB, FDL, and PM acknowledge support by the Fondazione ICSC, Spoke 3 Astrophysics and Cosmos Observations. National Recovery and Resilience Plan (Piano Nazionale di Ripresa e Resilienza, PNRR) Project ID CN{\_}00000013 ``Italian Research Center on  High-Performance Computing, Big Data and Quantum Computing"  funded by MUR Missione 4 Componente 2 Investimento 1.4: Potenziamento strutture di ricerca e creazione di "campioni nazionali di R\&S (M4C2-19 )" - Next Generation EU (NGEU), by INFN through the InDark initiative. DE acknowledges support from the Swiss National Science Foundation (SNSF) under grant agreement 200021{\_}212576. K.U. acknowledges support from the National Science and Technology Council of Taiwan (grant NSTC 112-2112-M-001-027-MY3) and the Academia Sinica Investigator Award (grant AS-IA-112-M04). This research was supported by Basic Science Research Program through the National Research Foundation of Korea (NRF) funded by the Ministry of Education (2019R1A6A1A10073887). This research was funded by the 2025 KAIST-U.S. Joint Research Collaboration Open Track Project for Early-Career Researchers, supported by the International Office at the Korea Advanced Institute of Science and Technology (KAIST).

\end{acknowledgements}

\bibliographystyle{aa}
\bibliography{ref}

\appendix
\section{Updates Relative to K24}
\label{sec:appen:K24_compare}
The fitting pipeline used in this analysis expands upon the one initially created by K24. During the process of integrating the WL analysis into the established pipeline and performing cross checks of our results with similar analyses, namely the one by \citet{Loris2025}, several updates were made to the algorithms described in K24. In addition, we have also utilized different observational data for some of the probes. Here we address each of the changes and discuss how they impact the fit. In several cases, we use the value of $\mathcal{R}_{LP}$ to quantify the overall impact of the update on our analysis relative to that of K24. When considering all of the updates, the result is that $\mathcal{R}_{LP}$ is approximately unchanged, and so we obtain overall consistent results with K24 as detailed in Section~\ref{sec:Results}.

\begin{itemize}
    \item \textbf{ACT DR6 Data Release}
\end{itemize}
Subsequent to the analysis of K24, the ACT DR6 maps were made publicly available \citep{Coulton2024,naess2025_ACTDR6}. The ACT DR6 maps are deeper than the DR4 maps utilized by K24, which results in more precise constraints on the electron pressure profile in our analysis. Consequently, as noted in Section~\ref{sec:Results}, the SZ data dominate over the X-ray temperature data in constraining the shape of the pressure profile. With this new data release, we also updated the method used to calculate the covariance matrix, with the primary change being a masking of map regions that contain ACT cluster detections from \citet{Hilton2021}. This results in a more accurate estimate of the (now lower) noise in the SZ Compton-$y$ map.

\begin{itemize}
    \item \textbf{Projected One-Dimensional X-ray SB}
\end{itemize}
While we adopt an identical approach to K24 in utilizing a projected one-dimensional X-ray SB profile at large radii, we obtain this profile in a different manner. Specifically, we utilize the azimuthal-median profiles calculated by \citet{Bartalucci2023}. In contrast, K24 computed the projected profile from the azimuthal mean of the two-dimensional SB map.
In addition, we discovered that the calculation of the associated uncertainties on this profile in K24 was incorrect, and had been inflated by an amount that approximately corresponds to the square root of the number of map pixels within each radial bin. As a result, the large-radius X-ray SB data were effectively down-weighted in the fit of K24, due to error estimates that were inflated by 1--2 orders of magnitude, and this resulted in a fit that was systematically lower than the observed data at those radii. The difference occurs at radii $\gtrsim 3.5$\arcmin, increasing to a factor of 2 at 5\arcmin and a factor of 10 at the maximum radius 7.4\arcmin. In aggregate, we find that the higher electron density recovered in our fit compared to that of K24 results in the value of $\mathcal{R}_{LP}$ decreasing by approximately 0.05.

\begin{itemize}
    \item \textbf{Projected One-Dimensional X-ray Temperature}
\end{itemize}

In their fit, K24 utilized the two-dimensional X-ray temperature map produced by \citet{Lovisari2024}. As part of our testing, we compared the results obtained from this approach to those obtained from the projected one-dimensional temperature profiles from \citet{Rossetti2024}, finding fitted temperature values approximately 10~per cent higher from the one-dimensional profiles. As a result, this change reduces the value of $\mathcal{R}_{LP}$ found in our analysis relative to that of K24 by approximately 0.20. We attribute this to the different weighting employed in the POS. In the case of the two-dimensional maps, the POS weighting is performed by the $\chi^2$ statistic used in the fit, and so it is an inverse-variance weighting. In contrast, the one-dimensional projected profiles are obtained from X-ray spectral fits to all the photons within a given annulus, which implies a weighting more similar to that suggested by \citet{Mazzotta2004}. As shown by the comparisons in \citet{Lovisari2024}, per-cluster differences comparable to the 10~per cent value we obtain for Abell 1689 are expected for these different weighting choices. Lacking a definitive test to establish which approach is least biased, we chose to use the projected one-dimensional profiles for this analysis, since those are the default temperature products produced within the CHEX-MATE collaboration. This multiplicative difference in temperature primarily impacts the fitted LOS elongation, as quantified below when describing the change in APEC normalization.

\begin{itemize}
    \item \textbf{Logrithmically-Spaced LOS Projection Integral}
\end{itemize}
In the development of the WL analysis, we found that the central values of the shear maps were unphysical when using a linear spacing for the numerical LOS integration of the three-dimensional elliptical coordinate. This was a result of the projection integration in Eq. (\ref{eq:Proj}), which was not sufficiently sampled at small radii. To remedy this, we updated the numerical integration to utilize log-spaced steps, with Eq. (\ref{eq:Proj}) becoming
\begin{equation}
F_{2D}(x_\xi;l_p,p_i) = 2l_pe_\parallel \text{ln}(10) \int_{x_\xi}^{\infty} F_{3D}(10^{x_\zeta};l_s,p_i) \frac{10^{2x_\zeta}}{\sqrt{10^{2x_\zeta} - x_{\xi}^{2}}} dx_\zeta.
\end{equation}
This change was also propagated to the SZ and X-ray SB projection integrals. The X-ray temperature calculation does not employ the log-spaced integral, since its value is far more uniform as a function of radius. Cross-checks were performed to ensure the log-spaced numerical integrals produce results consistent with the original linear-spaced numerical integrals for the gas quantities. Using a logarithmic basis decreased the integration time by three orders of magnitude for the SZ projection and by an order of magnitude for the X-ray SB projection relative to the linear basis. The fitted parameter values are not noticeably impacted by this update.

\begin{itemize}
    \item \textbf{Tolerance of the Numerical LOS Projection Integral}
\end{itemize}
To perform the numerical integration in Eq. \ref{eq:Proj}, we use \verb|scipy.integrate.quad|. This function allows for adjustment of the absolute and relative error tolerances. We found that when using the DR6 SZ map, to produce stable model maps the absolute and relative error tolerances needed to be adjusted from their default values of $1.49\times10^{-8}$ to $10^{-12}$. Stability was defined as model maps that vary by less than 0.01~per cent when the tolerances were lowered by an order of magnitude. We tested the model maps corresponding to the other observables and found the default tolerances were sufficient for them, so, to conserve computation time, the tolerance adjustment was only applied to the numerical projection integral corresponding to the SZ signal. This adjustment results in an overall decrease in the value of $\mathcal{R}_{LP}$ by 0.03.

\begin{itemize}
    \item \textbf{\xmm\ Vignetting Correction}
\end{itemize}
When constructing the projected X-ray SB maps and profiles from the triaxial model, K24 inadvertently applied a vignetting to mimic the \xmm\ response prior to comparing with the observed data. However, since the observed X-ray SB data have already been corrected for vignetting, this should not have been done, and it has now been removed from the analysis. In addition, the radial scaling of the vignetting correction was not correctly applied in K24. In aggregate, our updates to the vignetting correction result in the value of $\mathcal{R}_{LP}$ increasing by 0.02.

\begin{itemize}
    \item \textbf{APEC Normalization}
\end{itemize}
In K24, an additional factor of $(1+z)^{-1}$ was mistakenly included when using the Python package \verb|pyproffit| to calculate the emissivity. This error occurred because the package expects an input X-ray SB in units of the APEC normalization, which scales as $(1+z)^{-2}$, rather than the scaling of $(1+z)^{-3}$ from Eq.~\ref{eq:SB}. Thus, the normalization in K24 was mis-estimated by a factor of 1.18. Correcting for this results in the value of $\mathcal{R}_{LP}$ increasing by 0.18 in our analysis relative to K24. 

\begin{itemize}
    \item \textbf{XMM Response File}
\end{itemize}
As described in Section~\ref{sec:SZX_obs}, to calculate the emissivity, the \xmm\ instrument response within the chosen energy band must be accounted for. K24 inadvertently utilized the incorrect ARF for these data. For the analysis in this paper, we instead utilized an approximately universal \xmm\ ARF that is expected to be accurate for the observations of the CHEX-MATE clusters at the per-cent level. These two ARFs differ by approximately 7~per cent, which results in our fitted value of $\mathcal{R}_{LP}$ being larger by 0.07 relative to K24.

\section{Example Posteriors for the Model Fit of a Mock Cluster}

\begin{figure*}
    \centering
    \includegraphics[width=0.9\linewidth]{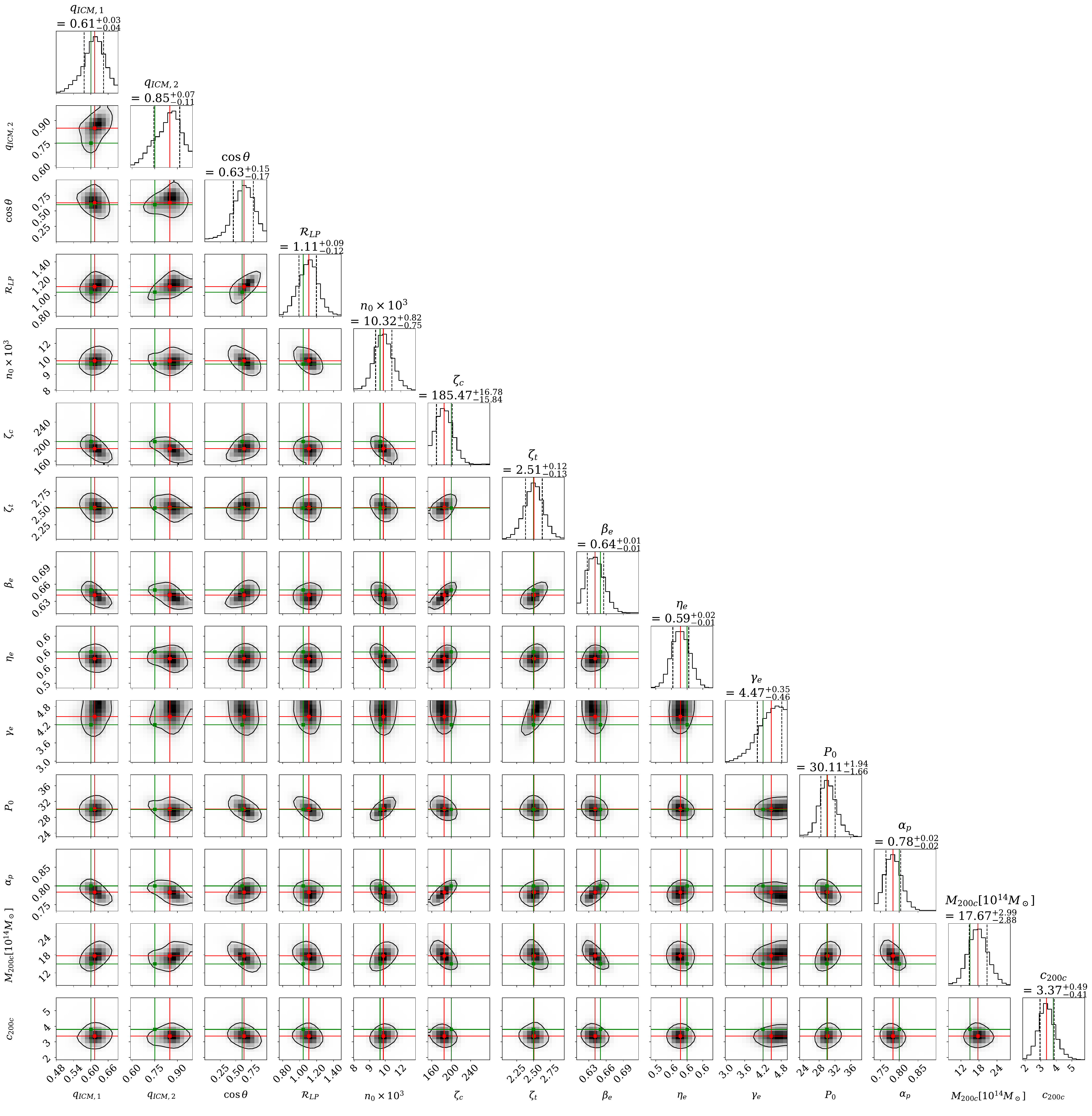}
    \caption{Posterior distributions estimated from our MCMC for a mock observation generated from a smooth model. The green vertical lines in each plot indicate the input parameters used to generate the mock observation maps and the red vertical lines represent the median value from the accepted MCMC samples. The value displayed above each histogram shows the median of the distribution, along with the $1\sigma$ (68~per cent) credible region, which is indicated by the dashed vertical lines in every plot. The black line in the two-dimensional distributions encloses the 68~per cent credible region.}
    \label{fig:toymodel_param_results}
\end{figure*}

\section{Posteriors for the Model Fit of the Observed Data}

\begin{figure*}
    \centering
    \includegraphics[width=0.9\linewidth]{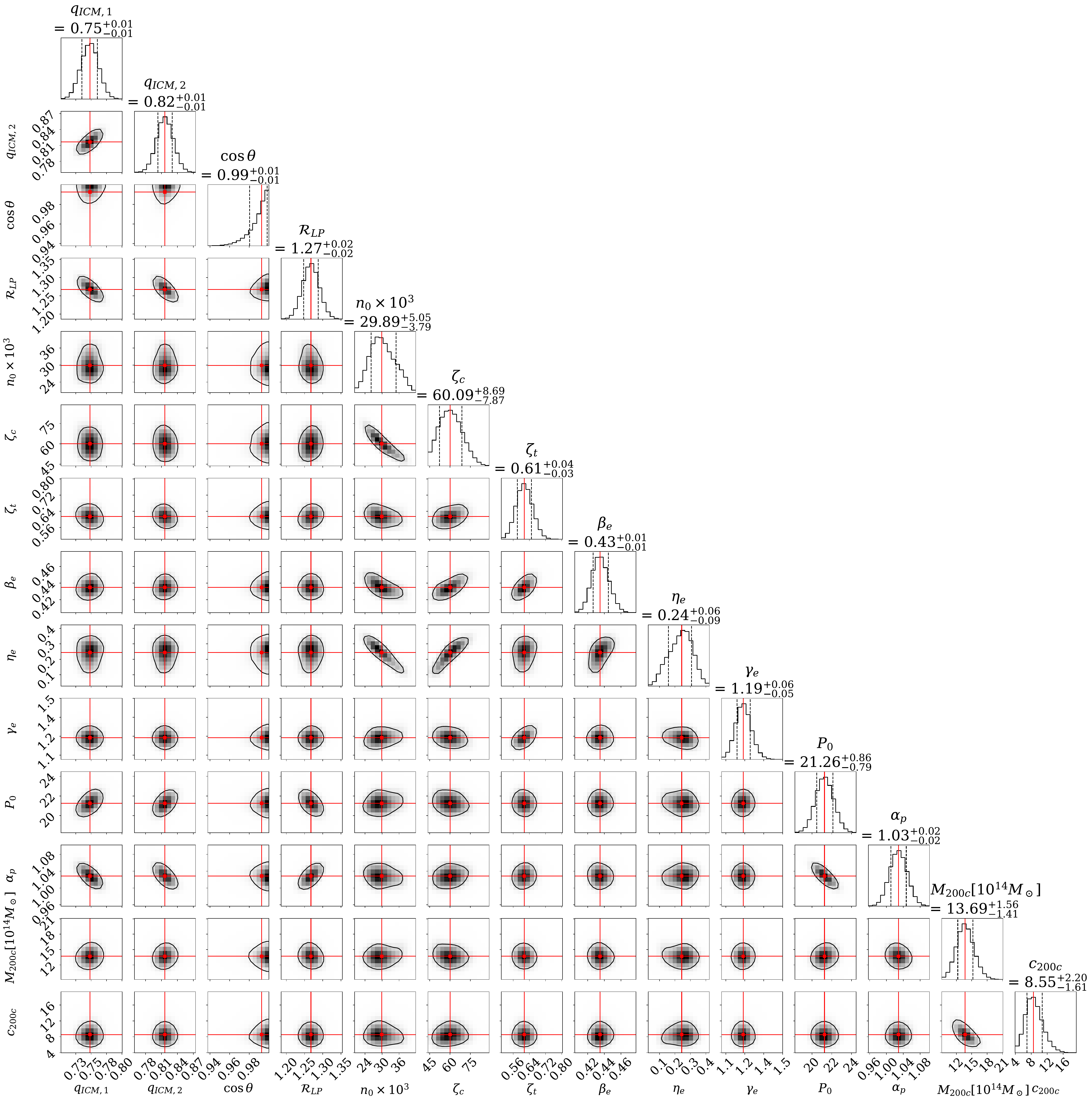}
    \caption{Posterior distributions of the model fit parameters for Abell 1689. The convention is identical to Fig.~\ref{fig:toymodel_param_results}.}
    \label{fig:A1689_cornerplot}
\end{figure*}

\section{Comparison With ICM Thermodynamic Profiles from K24}
\begin{figure*}
    \centering
    \includegraphics[width=0.45\linewidth]{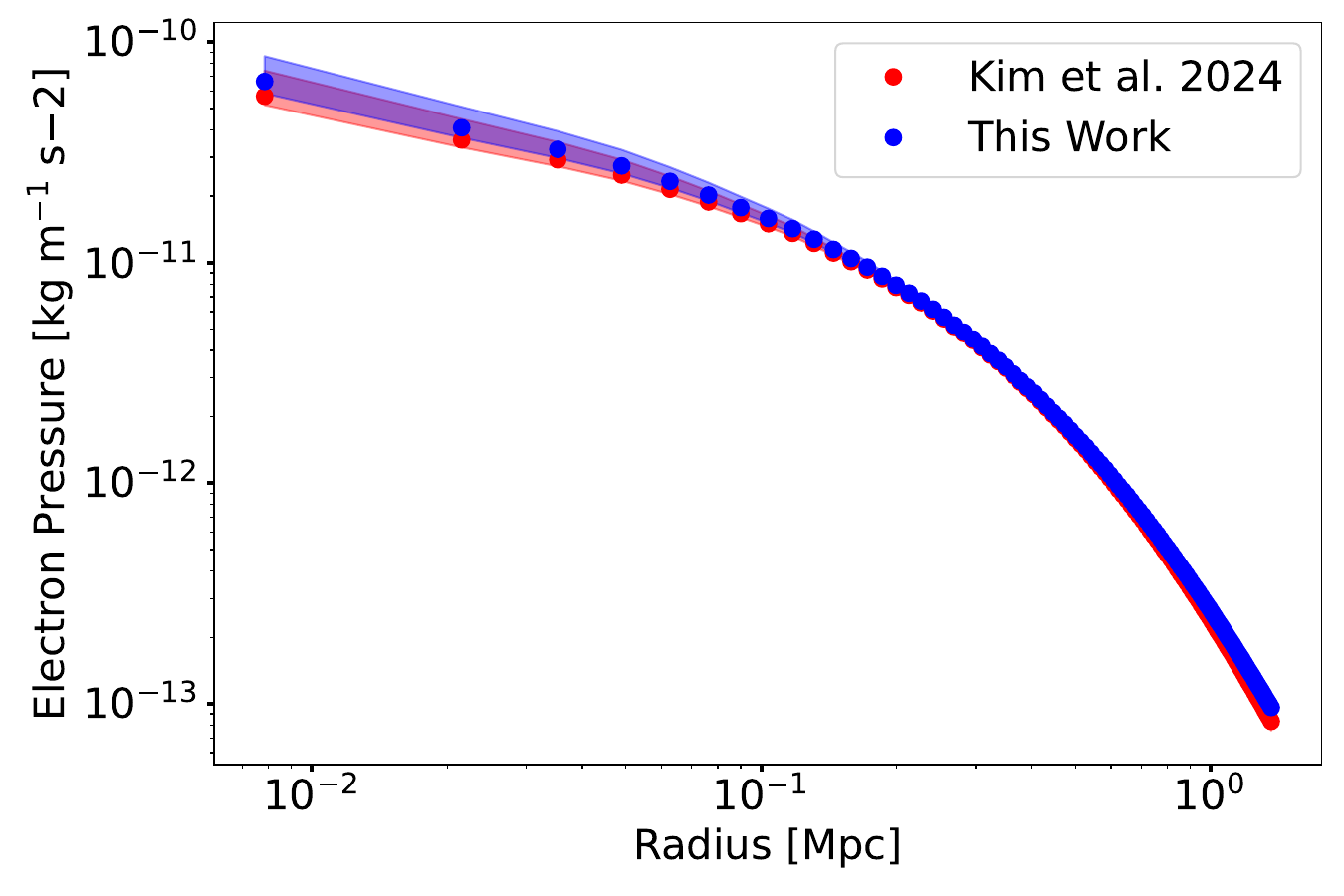}
    \includegraphics[width=0.45\linewidth]{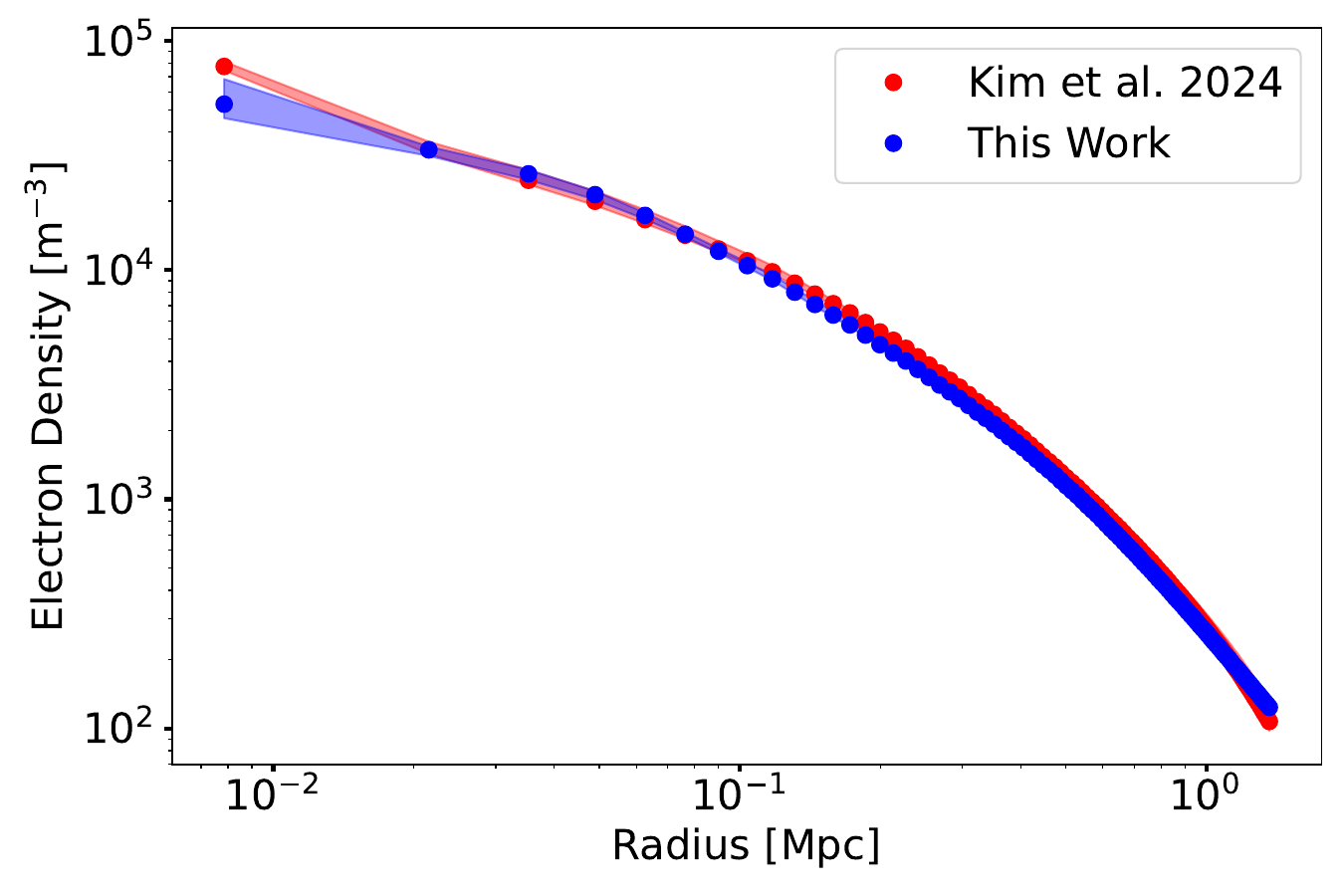}
    \caption{Electron density (left) and electron pressure (right) profiles generated from the accepted MCMC samples for the fit performed in this work (blue) and K24 (red).}
    \label{fig:eprofiles_compare}
\end{figure*}

\end{document}